\documentclass[aps,prc]{revtex4} 

\usepackage{multirow}  
\usepackage{dcolumn} 
\usepackage{amsfonts} 
\usepackage{amssymb}
\usepackage{latexsym}  
\usepackage{epsfig}
\usepackage{color}
\usepackage{lscape}

\newcommand{\be}{\begin{equation}}
\newcommand{\ee}{\end{equation}}
\newcommand{\ba}{\begin{array}}
\newcommand{\ea}{\end{array}}
\newcommand{\bea}{\begin{eqnarray}}
\newcommand{\eea}{\end{eqnarray}}
\def\half{{1\over2}}
\def\Hb{\mathbb{H}}
\def\Rb{\mathbb{R}}
\def\Cb{\mathbb{C}}
\def\so{\mathfrak{so}}
\def\u{\mathfrak{u}}
\def\su{\mathfrak{su}} 
 
\newcommand{\hf}{\textstyle \frac{1}{2} \displaystyle}   

\begin{document}

\title{Vector coherent state theory of the generic representations of $\so(5)$ in an  
$\so(3)$ basis}

\author{P.S.\ Turner$^\dag$\footnote{Current address: \emph{Institute for Quantum Information Science, University of Calgary, Calgary, Alberta T2N 1N4, Canada}}, D.J.\ Rowe$^\dag$, and  J.\ Repka$^\S$}

\affiliation{
$^\dag$Department of Physics, University of Toronto, Toronto,  Ontario M5S 1A7, Canada\\ 
$^\S$Department of Mathematics, University of Toronto, Toronto, Ontario M5S 2E4,
Canada }

\date{09 November 2005} 

\begin{abstract}
 
For applications of group theory in quantum mechanics, 
one generally needs explicit matrix representations of the spectrum generating algebras 
that arise in bases that reduce the symmetry group of some 
Hamiltonian of interest.
Here we use vector coherent state techniques to develop an algorithm 
for constructing the matrices for arbitrary finite-dimensional irreps 
of the SO(5) Lie algebra in an SO(3) basis.
The SO(3) subgroup of SO(5) is defined by regarding SO(5) as linear 
transformations of the five-dimensional space  of an SO(3) irrep of 
angular momentum two.
A need for such irreps arises in the nuclear collective model of 
quadrupole vibrations and rotations.
The algorithm has been implemented in MAPLE,  and 
some tables of results are presented.  

\end{abstract}

\maketitle

\section{Introduction} 

A vector coherent state (VCS) representation is a representation of a
group (or Lie algebra) on a space of vector-valued functions.
It is a representation induced from a multi-dimensional representation
of a subgroup.
Such representations have been used widely in the construction of explicit
representations of Lie algebras and Lie groups \cite{VCS,HLeBR,RLeBH,LeBR88,LeBR89,RR91},
in the construction of  shift tensors \cite{RR95}, and for the computation of
Clebsch-Gordan coefficients for reducing tensor product representations
\cite{RR97,RB00}.

The VCS construction for a representation of a group $G$  
involves two subgroups which play quite 
different roles.  A so-called  
``intrinsic'' subgroup (sometimes called the ``core" subgroup) acts in a known
way on a subspace of the representation of interest.  A second ``orbiter'' subgroup
 acts upon this subspace to generate the larger representation of the group $G$.  
A prototypical example of the construction is that for the dynamical group of the
rigid    rotor given by the semi-direct product
$G = \Rb^5\ltimes$SO(3) of an intrinsic $\Rb^5$
subgroup, which describes the  quadrupole moments (hence the shape) of an object,   and an 
orbiter group SO(3), corresponding to physical rotations of the object, which
describes its possible orientations.
The quantum mechanics of such a rotor are then described by the unitary irreducible
representations (irreps) of $G$.  

The key requirement is that the Lie algebras of the
intrinsic and orbiter groups,  together with those
elements of the complexified Lie algebra ${\frak g}_\Cb$
of $G$ which leave the intrinsic space invariant,  span the
complex extension of the Lie algebra.  Finding such groups is
often easier in the complex extension of $G$.
In many cases there are then  mathematically naturally choices of intrinsic and orbiter
subgroups for which the VCS construction of an induced representation is straightforward.
Unfortunately a mathematically natural choice often produces a  
representation in a basis which is not adapted to the symmetries of a physical problem.
A goal of this paper is to show how to construct representations of $\so(5)$ in a
basis which reduces a physically relevant $\so(3)\subset \so(5)$ subalgebra.

The group SO(5) and its Lie algebra $\so(5)$ arise in many physical contexts.
For example, they are needed for the classification of states in the Bohr-Mottelson model
\cite{BMbook} and
Interacting Boson model \cite{IBM} of nuclear collective states. 
They arise in a charge-independent pairing theory and in the use of isospin for the
classification of nuclear shell model basis states \cite{SMpairing,Hecht65}.
They have also been used for the study of algebraic many-body equations of motion methods
\cite{Klein} and high-temperature superconductivity \cite{HTc}.
Depending on the context, $\so(5)$ irreps may be required in an $\su(2)$ (e.g.,\ isospin)
or an $\so(3)$ (angular momentum) basis.
The isospin $\su(2)$ algebra is embedded in $\so(5)$ as a subalgebra 
$\su(2) \subset \su(2) \times \su(2)\cong \so(4) \subset \so(5)$.
 Thus, the required $\so(5) \supset \su(2)$ irreps are given in a basis that reduces the
Gel'fand chain  SO$(5)\supset$ SO$(4)\supset$ SO$(3) \supset$ SO$(2)$ \cite{Gel}. 
Such irreps were constructed years ago \cite{Hecht65,Kemmer} and
reconstructed more simply by VCS methods in Refs.\ \cite{HE,RLeBH}.
Note, however, that the  $\so(3) \subset \so(4)$ subalgebra of the
SO(3) group in the Gel'fand chain is not the same as the 
above-mentioned angular momentum algebra which generates a 
``geometrical'' SO(3) $\subset$ SO(5) subgroup of rotations of an associated
three-dimensional space; as a result the  construction of $\so(5)$ irreps in an
angular-momentum basis is more challenging.

The so-called one-rowed representations that occur in the decomposition of the Hilbert
space of the 5-dimensional harmonic oscillator
can be inferred in an $\so(3)$ basis from
the results of Cha\'con, Moshinsky and others
\cite{CMS} or from the SO(5) hyperspherical harmonics and Clebsch-Gordan coefficients
given in Ref.\ \cite{RTR}.  
An explicit VCS (vector coherent state) construction of irreps with highest
weights of the type $(v,0)$ and $(0,f)$ (this notation 
is explained below) was given by Rowe and Hecht \cite{RH}.  

In this paper we give a systematic construction of the generic
$(v,f)$ irreps in an $\so(3)$ basis.
In addition to its obvious relevance to the representation theory of
$\so(5)$, the construction is a prototype for a relatively sophisticated
application of VCS theory.

\section{Vector coherent state representations}

{\em Vector coherent state (VCS)  theory\/} is a generalisation
of standard (scalar) coherent state theory \cite{SCS}.
It was introduced \cite{VCS} for
the purpose of providing an explicit systematic construction of the
irreducible unitary representations of the compact and non-compact
symplectic Lie algebras. Simplicity and efficiency were achieved in the
construction by making use of the already well-known   representation
theory of the unitary subalgebras. Important aspects of the theory were
also introduced independently by Deenen and Quesne in their {\em partial
coherent state representations}
\cite{DQ}. Subsequently, VCS has been used to construct representations of
a large number of Lie algebras, groups, and superalgebras (cf.\ Ref.\
\cite{RR91} for a review).
Early applications of VCS theory gave realisations of the so-called {\em
holomorphic representations}  (reviewed by Hecht \cite{Hechtbook}).
A more general class of VCS representations was later
used in the construction of $\su(3)$ irreps in an $\so(3)$ basis
\cite{su3so3}.  It has also been shown that VCS theory is 
compatible with the theory of induced representations
\cite{RR91} and the theory of geometric quantisation \cite{GQ}.
A more general perspective on the theory was given in Ref.\ \cite{RR02}.

The construction of the finite-dimensional
irreps of $\so(5)$ in an $\so(3)$ basis has much in common with the
construction of irreps of $\su(3)$ in an $\so(3)$  basis. 
The $\su(3)$ Lie algebra is spanned by the components of $\so(3)$
tensors of angular momentum $L=1$ and $L=2$, while the $\so(5)$ Lie algebra is
spanned by  $\so(3)$ tensors of angular momentum $L=1$ and $L=3$.
However, whereas for $\su(3)$ it was possible to use scalar coherent state wave functions,  
a special case of VCS functions,  it proves to be essential to use
vector-valued  wave functions for $\so(5)$. 

The VCS theory of $\su(3)$ relies on the fact that the carrier space for an
$\su(3)$ irrep is spanned by the set of states generated by 
SO(3) rotations of a highest weight state.
The carrier space for a generic irrep of $\so(5)$ is generated by 
SO(3) rotations of  a set of highest \emph{grade} states.

\subsection{Highest grade states for an $\so(5)$ irrep}

\begin{figure}[!htb]
\setlength{\unitlength}{1cm}
\begin{center}
\begin{picture}(5,4)(-1.7,-1.7)
\put(0,0){\vector(0,1){2}}
\put(0,0){\vector(1,1){2}}
\put(0,0){\vector(1,0){2}}
\put(0,0){\vector(1,-1){2}}
\put(0,0){\vector(0,-1){2}}
\put(0,0){\vector(-1,-1){2}}
\put(0,0){\vector(-1,0){2}}
\put(0,0){\vector(-1,1){2}}
\put(0.4,-0.3){$X_0,F_0$}
\put(2.1,0){$F_+$}
\put(2.1,-2.1){$T_{-}$}
\put(0,-2.1){$X_-$}
\put(-2.5,-2.1){$S_{-}$}
\put(-2.5,0){$F_-$}
\put(-2.5,2.1){$T_{+}$}
\put(0,2.1){$X_+$}
\put(2.1,2.1){$S_{+}$}
\end{picture}
\end{center}
\caption{Root diagram for the $\so(5)$ Lie algebra.}\label{fig:roots}
\end{figure}
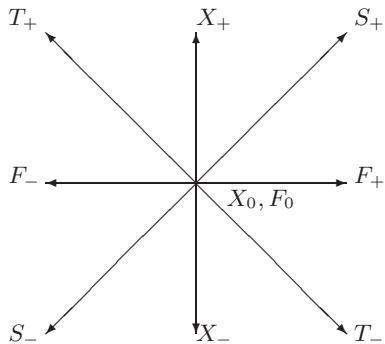

The $\so(5)$ Lie algebra is semisimple, of rank 2, and has 
the root diagram  shown in figure~\ref{fig:roots}.  
%\begin{figure}[t]%\vspace{0.5cm}
%  \epsfig{file=Roots.eps, height=1.5in} 
%  \caption{SO(5) root diagram
%  \label{fig:roots}} 
%\end{figure}
It is conventional to separate the roots of a semisimple Lie algebra into
positive and negative roots and to regard the corresponding root vectors as
raising and lowering operators, respectively.  Every irrep is then
characterised by a highest weight state.

Let $\alpha_1$ be the root corresponding to 
the vector $T_+$ and $\alpha_2$ be that corresponding to 
$F_+$.  
Then in the standard labelling scheme, the highest weight $\Lambda$ for an irrep is given by two integers
$\lambda_1$ and $\lambda_2$ such that $\Lambda = (\lambda_1 + \half\lambda_2)\alpha_1 + (\lambda_1 +
\lambda_2)\alpha_2$.  In keeping with the nuclear structure notation, we use the label $v=\lambda_1$ (the
``seniority''), and find it convenient to introduce the half-integer $f=\half\lambda_2$ since it also
labels a $\u(2)$ irrep, as is shown below.  Thus, we label an 
$\so(5)$ irrep $(vf)$; the highest weight is then given by $\Lambda = (v+f)(\alpha_1 + \alpha_2) + f\alpha_2$.

\begin{figure}[ht]
\setlength{\unitlength}{0.5cm}
\begin{center}
\begin{picture}(40,10)(-3,-5)
\put(5,0){\line(0,1){3}}
\put(5,0){\line(1,1){3}}
\put(5,0){\line(-1,1){3}}
\put(5,0){\line(1,0){3}}
\put(5,0){\line(-1,0){3}}
\put(5,0){\line(1,-1){3}}
\put(5,0){\line(0,-1){3}}
\put(5,0){\line(-1,-1){3}}
\put(1.5,-0.5){\dashbox{0.25}(7,1)}
\put(8.7,-0.1){\footnotesize{$\u(2)$ intrinsic (grade conserving)}}
\put(1.5,3.2){$T_{+}$}
\put(4.5,3.2){$X_+$}
\put(8.2,3.2){$S_{+}$}
\put(1,2.8){\dashbox{0.25}(8.4,1.2)}
\put(9.7,3.2){\footnotesize{grade raising}}
\put(1.5,-3.7){$S_{-}$}
\put(4.5,-3.7){$X_-$}
\put(8.2,-3.7){$T_{-}$}
\put(1,-4){\dashbox{0.25}(8.4,1.2)}
\put(9.7,-3.5){\footnotesize{grade lowering}}

\put(18,0){\circle*{0.2}}
\put(19,1){\circle*{0.2}}
\put(20,2){\circle*{0.2}}
\put(21,2){\circle*{0.2}}
\put(22,2){\circle*{0.2}}
\put(19,0){\circle*{0.2}}
\put(20,0){\circle*{0.2}}
\put(21,0){\circle*{0.2}}
\put(22,0){\circle*{0.2}}
\put(20,1){\circle*{0.2}}
\put(21,1){\circle*{0.2}}
\put(22,1){\circle*{0.2}}
\put(22,1){\circle*{0.2}}
\put(18,0){\line(1,1){2}}
\put(20,2){\line(1,0){4}}
\put(18,0){\line(-1,-1){1}}
\put(17,-1){\line(0,-1){1}}
\put(24,2){\circle*{0.2}}
\put(25,2){\circle*{0.2}}
\put(26,2){\circle*{0.2}}
\put(27,1){\circle*{0.2}}
\put(28,0){\circle*{0.2}}
\put(24,1){\circle*{0.2}}
\put(24,0){\circle*{0.2}} 
\put(25,1){\circle*{0.2}}
\put(25,0){\circle*{0.2}}
\put(26,0){\circle*{0.2}}
\put(26,1){\circle*{0.2}}
\put(27,0){\circle*{0.2}}
\put(24,2){\line(1,0){2}}
\put(26,2){\line(1,-1){2}}
\put(28,0){\line(1,-1){1}}
\put(29,-1){\line(0,-1){1}}
\put(19.5,1.5){\dashbox{0.25}(7,1)}
\put(15.8,2){\footnotesize{highest grade}}
\put(27,3){\vector(-1,-1){0.9}}
\put(27,3){\footnotesize{highest weight}} 
\put(27.8,2.2){$|(vf)f\rangle$}
\end{picture}
\end{center}
\caption{The $\u(2)$ intrinsic subalgebra and grade raising/lowering
operators.  The highest grade and highest weight of a generic $\so(5)$ irrep are
shown in the second diagram.}\label{fig:weights}
\end{figure}
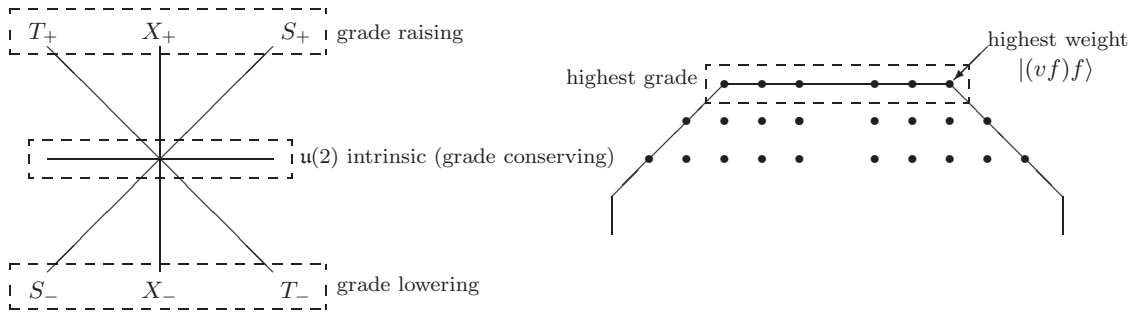

For present purposes, we
separate the root vectors into grade raising, grade conserving, and grade lowering
operators as shown in figure~\ref{fig:weights}. 
The horizontal grade-conserving root vectors 
$\{ F_\pm,F_0,X_0\}$ then
define what we shall refer to as an {\it intrinsic\/} or {\it core\/} $\u(2)$ subalgebra.
This grading of the $\so(5)$ Lie algebra generates a grading of any irrep.
Each irrep has a set of highest grade states
$\{|(vf)m\rangle\}$ that are annihilated by the grade-raising operators
$\hat T_+$, $\hat X_+$, and
$\hat S_+$, and carry an irrep of the above-mentioned intrinsic
$\u(2)$ algebra; the highest grade states satisfy the equations
\bea  
&\hat S_+|(vf)m\rangle
 = \hat X_+|(vf)m\rangle = \hat T_+|(vf)m\rangle =0 \,, & \label{eq:1}\\
&\hat X_0 |(vf)m\rangle
 = (v+f)|(vf)m\rangle \,,\quad \hat F_0 |(vf)m\rangle=m|(vf)m\rangle \,,&\\
&\hat F_\pm |(vf)m\rangle
 = \sqrt{(f\mp m)(f\pm m+1)}\, |(vf)m\pm 1\rangle \,.& \label{eq:3}
\eea

The weights for the highest grade states of a generic irrep of
$\so(5)$ are as illustrated in figure~\ref{fig:weights}.
%\begin{figure}[t]%\vspace{0.5cm}
%  \epsfig{file=Basics.eps, height=2.2in}  
%  \caption{Weight diagram for a generic irrep of $\so(5)$ showing the
%weights of the highest grade states.   
%  \label{fig:weights}} 
%\end{figure}
A set of wave functions $\{ \xi^{(vf)}_m; m= -f,\dots,f\}$ for these highest grade states are regarded
as {\em intrinsic\/} wave functions in the VCS construction --- it is in the Hilbert 
space of these intrinsic functions that the VCS wave functions take their {\em vector} values.

\subsection{Holomorphic VCS wave representations}

Let $|\psi\rangle$ be a state in the carrier space of an $\so(5)$ irrep $(vf)$.  
Then a holomorphic VCS wave function is defined for this
state by
\be 
\Psi(z)
 = \sum_m \xi^{(vf)}_m \langle (vf)m | e^{\hat z} | \psi\rangle,
\ee
where
\be 
\hat z = z_1 \hat S_+ + z_2\hat X_+ + z_3\hat T_+ 
\ee
and $z=(z_1,z_2,z_3)$ is a set of complex numbers.
The corresponding VCS representation $\Gamma$ of the $\so(5)$ Lie algebra
is defined by
\be 
[\Gamma(X)\Psi] (z) = \sum_m \xi^{(vf)}_m
\langle (vf)m | e^{\hat z} \hat X|\psi\rangle , \quad X\in\so(5) .
\ee
Such holomorphic representations are natural generalisations of the
familiar Bargmann-Segal representations \cite{BS} of the Heisenberg-Weyl
algebras.  They were the first to be considered in the formulation of VCS
theory \cite{VCS}.  However, in practical applications they are not always the most
useful.   In particular, they do not reduce the $\so(3)\subset\so(5)$ angular
momentum subalgebra.

\subsection{VCS wave functions in an SO(3)-coupled basis}

The group SO(3) can be embedded as a subgroup in SO(5) in many ways.
We consider the SO(3) subgroup defined 
up to conjugation
by regarding SO(5) as a group of
orthogonal transformations of the five-dimensional carrier space for an
$L=2$ irrep of SO(3).  This embedding is motivated by the rotational properties 
of the five quadrupole degrees of freedom in the nuclear collective model.
The construction of an SO(3)-coupled basis for a VCS irrep of $\so(5)$
then parallels a similar construction of an SO(3)-coupled basis for a VCS irrep 
of $\su(3)$ \cite{su3so3}. 
The $\so(5)$ construction makes use of the following theorem,   
which constrains the choice of the SO(3) subgroup. 
\medskip
\noindent 

{\em Theorem 1:} 
Provided no $\so(3)$ angular momentum
operator lies within the $\u(2)$ intrinsic subalgebra, the set of
states $\{ \hat R(\Omega) |(vf)m\rangle ;\, m=-f,\ldots, f;
\Omega \in$ SO$(3)\}$ obtained by all SO(3) rotations of an orthonormal basis for the 
highest grade subspace spans the Hilbert space for the $\so(5)$ irrep
$(vf)$.

\medskip
\noindent

{\em Proof:}
The set of states generated by  repeated application of the lowering
operators $\{ \hat S_-,\hat X_-, \hat T_-\}$ to the highest grade states
spans the Hilbert space of the irrep. Now, if $\{\hat{L}_i;\:i=1,2,3\}$ is
a hermitian basis for the $\so(3)
\subset \so(5)$ subalgebra, then each $\hat L_i$ can be expanded
$\hat{L}_i = \hat{L}_i^- + \hat{L}_i^0 + \hat{L}_i^+$, where $\hat{L}_i^-$
is a grade lowering operator, $\hat{L}_i^0$ is of grade zero, and
$\hat{L}_i^+$ is a grade raising operator. By hermiticity, if $\hat{L}_i$ 
has a non-zero component  $\hat{L}_i^+$,  it must also have a non-zero
$\hat{L}_i^-$ component.  Thus, if no $\hat{L}_i$ lies in the zero grade
$\u(2)$ subalgebra, then each $\hat{L}_i$ must have a nonzero
$\hat{L}_i^-$ component.  By linear independence, it must be that the
span of $\{\hat{L}_i^-\}$ equals the span of
$\{\hat S_-,\hat X_-,\hat T_-\}$.   \hfill \textbf{QED}

\medskip

This theorem means that an arbitrary state $|\psi\rangle$ in an irrep
$(vf)$ is defined by the set of overlaps $\{ \langle (vf)m|\hat
R(\Omega)|\psi\rangle ; m = -f,\dots , +f; \Omega \in$ SO$(3)\}$, 
provided that the SO(3) subgroup is chosen as required by the theorem, 
which we assume from now on.
It also means that, if $\{ |(vf)\tau LM\rangle\}$ is an 
SO$(3)$-coupled basis for an $\so(5)$ irrep and  $\{ |(vf)m\rangle\}$,
with wave functions  $\{\xi^{(vf)}_m\}$, is
an orthonormal $\u(2)$ basis of highest grade states for this irrep,  then
the basis states $\{ |(vf)\tau LM\rangle\}$ have VCS wave functions
given as vector-valued functions over SO(3) by
\be 
\Phi^{(vf)}_{\tau LM}(\Omega) = \sum_m \xi^{(vf)}_m \langle (vf)m|\hat
R(\Omega) |(vf)\tau LM\rangle \,. \label{eq:7}
\ee
These wave functions are very much like rotor-model wave functions 
\cite{BMbook}.
Indeed, with basis states chosen to have good SO(3) transformation
properties, they can be expanded
\be 
\Phi^{(vf)}_{\tau LM}(\Omega) = \sum_{mK} \xi^{(vf)}_m \langle (vf)m|
(vf)\tau LK\rangle \, \mathcal{D}^L_{KM}(\Omega) \,,\label{eq:8}
\ee
where $\mathcal{D}^L(\Omega)$ is a Wigner rotation matrix. 
It follows that a basis state $|(vf)\tau LM\rangle$ is characterised by
the set of expansion coefficients
\be  
b^{(vf)}_{mK}(\tau L) = \langle (vf)m| (vf)\tau LK\rangle \,.
\ee

The following gives a systematic procedure for determining these
coefficients and for deriving the transformations of these coefficients by
elements of the $\so(5)$ Lie algebra as defined by the VCS representation
\be  
[\Gamma(X) \Phi^{(vf)} _{\tau LM} ] (\Omega) =
\sum_m \xi^{(vf)}_m \langle (vf)m|\hat
R(\Omega) \hat X|(vf)\tau LM\rangle , \quad X\in \so(5) .
\label{eq:10}
\ee

\section{Representation spaces for $\so(5)$}

\subsection{A subspace of harmonic oscillator states}

\begin{figure}[ht]
\setlength{\unitlength}{1cm}
\begin{center}
\begin{picture}(2,3)
\put(0,1){\circle*{0.1}}
\put(1,2){\circle*{0.1}}
\put(1,1){\circle*{0.1}}
\put(1,0){\circle*{0.1}}
\put(2,1){\circle*{0.1}}
\put(0,1){\line(1,1){1}}
\put(0,1){\line(1,-1){1}}
\put(1,0){\line(0,1){2}}
\put(0,1){\line(1,0){2}}
\put(1,2){\line(1,-1){1}}
\put(1,0){\line(1,1){1}}
\put(1.1,2){$|(10)0\rangle$}
\put(0.7,-0.5){$(10)$}
\end{picture}
\qquad \qquad \qquad \qquad \qquad \qquad
\begin{picture}(1,1)
\put(0,0.5){\line(0,1){1}}
\put(0,1.5){\line(1,0){1}}
\put(1,0.5){\line(0,1){1}}
\put(0,0.5){\line(1,0){1}}
\put(0,0.5){\circle*{0.1}}
\put(1,0.5){\circle*{0.1}}
\put(1,1.5){\circle*{0.1}}
\put(0,1.5){\circle*{0.1}}
\put(1.1,1.5){$|(0\half)\half\rangle$}
\put(-1.6,1.5){$|(0\half)\textstyle\frac{-1}{2}\rangle$}
\put(0.2,-0.5){$(0\half)$}
\end{picture}
\end{center}
\caption{The weight diagrams for the fundamental irreps $(10)$ and $(0\half)$ with highest 
grade states labelled $|(vf)m\rangle$.\label{fig:basicweights}}
\end{figure}
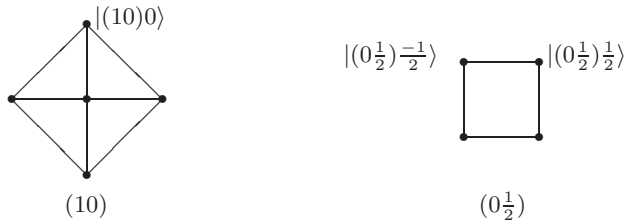

Irreps of $\so(5)$ can be built up from its two fundamental irreps with
highest weights $(10)$ and $(0\hf)$. 
The former is the fundamental five dimensional irrep, 
the latter is the fundamental four dimensional irrep.
Both weight diagrams are shown in figure~\ref{fig:basicweights}.
These fundamental irreps are carried by the spaces generated by 
the raising operators 
$\{\eta^{\dag}_{\nu};\nu= 0, \pm 1,\pm 2\}$ 
of a five-dimensional harmonic oscillator with symmetry group U(5)
and by the raising operators 
$\{\zeta^{\dag}_{m}; m=\pm \frac{1}{2},\pm \frac{3}{2}\}$ 
of a four-dimensional harmonic oscillator with symmetry group U(4), respectively.  The operators
$\{\eta^\dagger_\nu\}$ and $\{ \zeta^\dagger_m\}$, together with the
corresponding lowering operators, satisfy the usual boson commutation
relations
\bea  
& [\eta^\mu, \eta^\dagger_\nu]
 = \delta_{\mu\nu} \hat I \,, \quad [\eta^\mu, \eta^\nu]
 = [\eta^\dagger_\mu, \eta^\dagger_\nu]=0 \,, &\\
& {[}\zeta^m, \zeta^\dagger_n]
 = \delta_{mn} \hat I \,, \quad [\zeta^m, \zeta^n]
 = [\zeta^\dagger_m, \zeta^\dagger_n]=0 \,,&
\eea
where we use the notation $\eta^\nu = (\eta^\dag_\nu)^\dag$ and
$\zeta^m =(\zeta^\dag_m)^\dag$.  

The invariants of U(5)
and U(4) are given by their respective number operators
\be
\hat n_\eta
 = \eta^\dag \cdot \eta
 = \sum_\nu \eta^\dag_\nu \eta^\nu, 
\quad
\hat n_\zeta
 = \zeta^\dag \cdot \zeta
 = \sum_m \zeta^\dag_m \zeta^m \, . 
\ee
For the natural SO(3) embedding (defined to within conjugation), we can
regard these number operators as coupled products, e.g., 
\be 
\hat n_\zeta
 = 2\sum_m (\textstyle\frac{3}{2},-m,\textstyle\frac{3}{2},m| 0,0) \zeta^\dag_m \zeta_{-m}
 = {\displaystyle \sum_m} (-1)^{\frac{3}{2}+m} \zeta^\dag_m \zeta_{-m} \, ,
\ee
where
$$\zeta_{-m} = (-1)^{\frac{3}{2}+m} \zeta^{m} .$$
Similarly, we can define
\be
\eta_{-\nu} = (-1)^{\nu} \eta^{\nu} .
\ee 

The fundamental five-dimensional irrep of the group SO(5) can be realised
as the group of special orthogonal transformations of the creation
operators
$\{ \eta^\dagger_\nu ; \nu = 0 , \pm 1,\pm 2\}$ 
that leave invariant the quantity
$\eta^\dagger\cdot \eta^\dagger =\sum_\nu (-1)^\nu \eta^\dagger_\nu
\eta^\dagger_{-\nu}$.  This realisation exhibits SO(5) as a
subgroup of  U(5). The four-dimensional fundamental irrep is a spinor irrep of
SO(5). It can be realised as the group of special orthogonal
transformations of  the (boson) creation operators
$\{ \zeta^\dagger_m ; m = \pm\hf , \pm \textstyle\frac{3}{2}\}$ 
that leave invariant the quantity
$\zeta^\dagger\cdot \zeta^\dagger=\sum_m (-1)^{\frac{3}{2} +m}
\zeta^\dagger_m
\zeta^\dagger_{-m}$. This realisation exhibits 
USp(4), the two-fold cover of SO(5),  
as a subgroup of U(4).
Because every irrep of a group is contained in a tensor product 
of copies of  
its fundamental irreps, it follows that every irrep of
$\so(5)$ can be realised  on a subspace of the tensor product $\Hb
=\Hb^{(5)}\otimes\Hb^{(4)}$, where
$\Hb^{(n)}$ is the Hilbert space of an $n$-dimensional harmonic oscillator.
Highest grade states for an $\so(5)$ irrep in $\Hb$ are given by 
\be 
|(vf)m\rangle = \frac{
(\zeta^\dagger_{3/2})^{f+m}(\zeta^\dagger_{1/2})^{f-m} (\eta^\dagger_2)^v}
{\sqrt{
(f+m)!(f-m)!v!}}|0\rangle \,, \quad m = -f, \dots  , f \,. \label{eq:HiGradeStates}
\ee

\subsection{A model space of VCS wave functions}
 
A model space for a compact Lie group $G$ is a representation of $G$ which  is a
direct sum of irreps comprising precisely one copy from each equivalence class of irreps
\cite{BBG}.
As emphasised by Biedenharn and Flath \cite{BF}, a model space provides a valuable
framework for a realisation of the tensor algebra of the group.
We now show that a Hilbert space of VCS wave functions for
the states of
$\Hb$ provides a model space with very useful properties.

It follows from Eq.\ (\ref{eq:7}) that, provided the conditions of Theorem
1 are satisfied, any state
$|\psi\rangle$ in $\Hb$ has VCS wave function
$\Psi$ given by
\be 
\Psi(\Omega) = \sum_{vfm} \xi^{(vf)}_m \langle (vf)m|\hat
R(\Omega) |\psi\rangle .
\ee
Now observe that, if the wave functions $\{ \xi^{(vf)}_m\}$ of the highest
grade states are expressed in Bargmann form as the holomorphic functions
\be 
\xi^{(vf)}_m(x,y) = \frac{x_1^{f+m}x_2^{f-m}
y^v}{\sqrt{(f+m)!\,(f-m)!\, v!}} ,  \label{eq:15}
\ee
then, using Eq.\ (\ref{eq:HiGradeStates}),  the $\u(2)$-intertwining operator
$\sum_{vfm} \xi^{(vf)}_m \langle (vf)m|$
can be expressed
\be  
\sum_{vfm} \xi^{(vf)}_m \langle (vf)m| = \langle 0 |e^{\hat \mathfrak{X}} ,
\ee
where 
\be 
\hat \mathfrak{X} = x_1 \zeta^{3/2}+ x_2 \zeta^{1/2} + y\eta^2 .
\ee
Thus, any state $|\psi\rangle$ in $\Hb$ has VCS wave function defined by
\be 
\Psi(\Omega) =\langle 0 |e^{\hat \mathfrak{X}}\hat R(\Omega) |\psi\rangle .
\label{eq:18} 
\ee

Let $\mathcal{F}^{(vf)}$ denote the Hilbert space of VCS wave functions
for an $\so(5)$ irrep of highest weight $(vf)$, relative to the inner
product inherited from that of $\Hb$. The  space of all VCS wave
functions for states in
$\Hb$ is then the direct sum 
\be \mathcal{F} = \bigoplus_{(vf)} \mathcal{F}^{(vf)} ,
\ee
of all such Hilbert spaces.  Hence, by construction, it is a model space for
$\so(5)$.  The following theorem, 
which generalises a theorem of Rowe and Hecht \cite{RH},
shows that $\mathcal{F}$ is also a ring
and that it can be generated from the VCS wave functions of the fundamental irreps
(1,0)  and (0,$\half$).

\medskip
\noindent 

{\em Theorem 2:} 
If $\Phi = \Phi_1\odot\Phi_2$ is the {\it model product\/} of VCS wave functions
defined by
\begin{equation}
  \Phi(x,y,\Omega) = \Phi_1(x,y,\Omega)\, \Phi_2(x,y,\Omega) \,,
\end{equation}
then, if $\Phi_1$ is in $\mathcal{F}^{(v_1f_1)}$ and 
$\Phi_2$ is in $\mathcal{F}^{(v_2f_2)}$, the product $\Phi$  is in 
 $\mathcal{F}^{(v_1+v_2,f_1+f_2)}$.
Moreover, if $\mathcal{F}^{(v_1f_1)}\odot\mathcal{F}^{(v_2f_2)}$ denotes
the linear span of model products of the functions of $\mathcal{F}^{(v_1f_1)}$
with the functions of $\mathcal{F}^{(v_2f_2)}$, then
\be 
\mathcal{F}^{(v_1f_1)}\odot\mathcal{F}^{(v_2f_2)}
= \mathcal{F}^{(v_1+v_2,f_1+f_2)}.
\ee

\medskip
\noindent 

{\em Proof:} 
Let $\hat Z_1^{(v_1f_1)}$ be a homogeneous polynomial of degree
$v_1$ in the $\{ \eta^\dagger_\nu\}$ raising operators and  degree
$2f_1$ in the $\{\zeta^\dagger_m\}$ raising operators  which creates a 
state with VCS wave function
\be  
\Phi^{(v_1f_1)}(\Omega) = \langle 0|  e^{\hat \mathfrak{X}} \hat
R(\Omega)
\hat Z_1^{(v_1f_1)} |0\rangle \,.
\ee
Let
\be 
\hat Z_1^{(v_1f_1)}(\Omega) =  \hat R(\Omega)\,
\hat Z_1^{(v_1f_1)}\, \hat R(\Omega^{-1}) \,,
\ee 
denote the corresponding rotated operator. 
Then
\bea  
\Phi^{(v_1f_1)}(\Omega) &=& \langle 0|  e^{\hat \mathfrak{X}}
\hat Z_1^{(v_1f_1)}(\Omega) e^{-\hat \mathfrak{X}}  |0\rangle \nonumber\\
&=& \langle 0|\Big( \hat Z_1^{(v_1f_1)}(\Omega) +
[\hat \mathfrak{X}, \hat Z_1^{(v_1f_1)}(\Omega)]  + 
{1\over 2!}[\hat \mathfrak{X},[\hat \mathfrak{X}, \hat Z_1^{(v_1f_1)}(\Omega)]] + \cdots \Big) 
|0\rangle\,.
\eea
The successive terms in the sequence of multiple commutators inside the
brackets are homogenous polynomials of decreasing degree in the raising
operators.
This sequence terminates with a polynomial of degree zero (a number) at
the $(v_1+2f_1)$'th term.
Moreover, this last term is the only term in the sequence with non-vanishing
vacuum expectation value.  It follows that
\be  
\Phi^{(v_1f_1)}(\Omega)
 = {1\over (v_1+2f_1)!} 
\underbrace{[\hat \mathfrak{X}, [\hat \mathfrak{X},[ \cdots\hat Z_1^{(v_1f_1)}(\Omega)]] \cdots
]}_{v_1\!+\!2f_1\; {\rm commutators}}\,. 
\ee
If $Z_2^{(v_2f_2)}$ is similarly defined for the wave function 
$\Phi^{(v_2f_2)}_2$, then the wave function $\Phi =
\Phi^{(v_1f_1)}_1\odot
\Phi^{(v_2f_2)}_2$ has values given by
\bea 
\Phi(\Omega)
&=& \langle 0| \Big(e^{\hat \mathfrak{X}}\hat Z_1^{(v_1f_1)}(\Omega) e^{-\hat
X}\Big)
\Big(e^{\hat \mathfrak{X}}\hat Z_2^{(v_2f_2)}(\Omega) e^{-\hat \mathfrak{X}}\Big) |0\rangle
\nonumber\\
 &=& \langle 0|  e^{\hat \mathfrak{X}} \hat R(\Omega)
\hat Z_1^{(v_1f_1)}\hat Z_2^{(v_2f_2)}|0\rangle \,.  
\eea 
The state $\hat Z_1^{(v_1f_1)}\hat Z_2^{(v_2f_2)}|0\rangle$ does
not necessarily belong to an irreducible $\so(5)$ subspace.
However, it is of degree $(v_1+v_2)$ and $2(f_1+f_2)$ in the
$\{ \eta^\dagger_\nu\}$ and $\{ \zeta^\dagger_m\}$ operators, respectively,
and, therefore, only the component of the  state $\hat
Z_1^{(v_1f_1)}
\hat Z_2^{(v_2 f_2)}|0\rangle$ that does lie within the irreducible SO(5) 
subspace of highest weight $(v_1+v_2,f_1+f_2)$ will have
non-zero overlap with $\langle 0|  e^{\hat \mathfrak{X}} \hat R(\Omega)$.
Thus, $\Phi$ is a VCS wave function for a state belonging to an SO(5) irrep
of highest weight $(v_1+v_2,f_1+f_2)$ as claimed by the
theorem. 
The second part of the theorem follows from the observation that, if
$\{ \hat Z_{\alpha}^{(vf)}\}$ is a basis for the linear space $P^{(vf)}$ of
homogeneous polynomials of degree $v$ in $\{\eta^\dag_\nu\}$ and degree $2f$
in $\{\zeta^\dag_m\}$, then the space $P^{(v_1+v_2,f_1+f_2)}$ is spanned
by the  products 
$\{ \hat Z_{\alpha_1}^{(v_1f_1)}\hat Z_{\alpha_2}^{(v_2f_2)}\}$.   
\hfill \textbf{QED}

\medskip

The theorem shows that $\mathcal{F}$ can be constructed from 
the two fundamental irreps, and that this process will generate 
all the irreps of $\so(5)$. 
The importance of a model space is 
that, since it is multiplicity free, calculations performed in 
$\mathcal{F}$ are not complicated by any need to keep track of  
equivalent copies of the same $\so(5)$ irreps.

\section{Bases for the $\so(5)$ Lie algebra}

The VCS representations of $\so(5)$ in an SO(3)-coupled basis make use of
two bases for the $\so(5)$ Lie algebra:
a Cartan basis of root vectors and a basis of components of SO(3) tensors.

\subsection{A Cartan basis}

Starting from two copies of the (10) irrep, one spanned by the creation
operators $\{\eta^{\dag}_{\nu}\}$ and one spanned by the annihilation
operators  $\{\eta_{\nu}\}$, one can construct a realisation of the 
the $\so(5)$ Lie algebra,  which carries a (01) irrep,  by taking an
antisymmetric tensor product of the two.  This gives the $\so(5)$ root vectors of figure\
\ref{fig:roots} as a subset of $\u(5)$ operators
\bea 
& \hat S_+ = \eta^\dag_1\eta_2 - \eta^\dag_2\eta_1 = 
\eta^\dag_1\eta^{-2} + \eta^\dag_2\eta^{-1}  \,,&\\
&\hat T_+ = \eta^\dag_{-1}\eta_2 - \eta^\dag_2\eta_{-1} = 
\eta^\dag_{-1}\eta^{-2} + \eta^\dag_2\eta^{1}  \,,&\\
&\hat X_+ = \sqrt{2}\, ( \eta^\dag_2\eta_0 - \eta^\dag_0 \eta_2)
=\sqrt{2}\, ( \eta^\dag_2\eta^0 - \eta^\dag_0 \eta^{-2}) \,,&\\
&\hat F_+ = \sqrt{2}\, ( \eta^\dag_1\eta_0 - \eta^\dag_0 \eta_1)
=\sqrt{2}\, ( \eta^\dag_1\eta^0 + \eta^\dag_0 \eta^{-1}) \,,& \\
&\hat S_- = (\hat S_+)^\dag \,,\quad \hat T_- = (\hat T_+)^\dag \,,&\\
&\hat X_- = (\hat X_+)^\dag \,, \quad \hat F_- = (\hat F_+)^\dag \, .&
\eea

A basis for the Cartan subalgebra is given by 
\bea 
&\hat X_0 = \half [\hat X_+, \hat X_-] =
\eta^\dag_2\eta^2-\eta^\dag_{-2}\eta^{-2} \,,&\\
& \hat F_0 = \half [\hat F_+, \hat F_-] =
\eta^\dag_1\eta^1-\eta^\dag_{-1}\eta^{-1} \,.&
\eea

A realisation of the $\so(5)$ algebra as a subalgebra of $\u(4)$ is
similarly obtained from the symmetric tensor product of two copies of the
$(0,\hf)$ irrep:
\bea 
& \hat S_+ = -\zeta^\dag_{3/2}\zeta^{-3/2}  \,, \quad 
  \hat T_+ =  \zeta^\dag_{1/2}\zeta^{-1/2}   \,,&\\ 
& \hat X_+ =  \zeta^\dag_{1/2} \zeta^{-3/2} - \zeta^\dag_{3/2}\zeta^{-1/2} \,,&\\ 
& \hat F_+ =\zeta^\dag_{3/2}\zeta^{1/2} + \zeta^\dag_{-1/2} \zeta^{-3/2} \,,&
\eea
\bea 
\hat X_0 =\textstyle \half\big(\zeta^\dag_{3/2}\zeta^{3/2} +\zeta^\dag_{1/2}  \zeta^{1/2}
- \zeta^\dag_{-1/2}\zeta^{-1/2} - \zeta^\dag_{-3/2}\zeta^{-3/2} )\,, \\
\hat F_0 =\textstyle \half\big(\zeta^\dag_{3/2}\zeta^{3/2} - \zeta^\dag_{1/2} 
\zeta^{1/2} + \zeta^\dag_{-1/2}\zeta^{-1/2} - \zeta^\dag_{-3/2}\zeta^{-3/2} )\,.
\eea

\subsection{An SO(3) tensor basis}

Let $\{\hat L_k\}$ denote a set of angular momentum operators for the
$\so(3)$ subalgebra of $\so(5)$ and let
$\{ d^\dag_M ; M = 0,\pm 1, \pm 2\}$ and 
$\{ p^\dag_M; M = \pm \half,\pm {3\over 2}\}$  denote linear combinations of the
$\{\eta^\dag_\nu\}$ and $\{ \zeta^\dag_m\}$ operators, respectively, 
which satisfy the commutation relations
\bea  
[\hat L_0, d^\dag_M] = Md^\dag_M, &\quad&
 [\hat L_\pm, d^\dag_M] = \sqrt{(2\mp M)(3\pm M)}\, d^\dag_{M\pm 1} , \\
{[}\hat L_0, p^\dag_M{]} = Mp^\dag_M, &\quad&
 {[}\hat L_\pm, p^\dag_M{]} = \textstyle \sqrt{(\frac{3}{2}\mp
M)(\frac{5}{2}\pm M)}\, p^\dag_{M\pm 1} .
\eea
A basis for a realisation of the $\so(5)$ Lie algebra on the combined
(tensor product) Hilbert spaces of the four- and five-dimensional harmonic
oscillators is then provided by the $(L=1)$- and $(L=3)$-coupled operators
\bea 
&\hat L_k = -\sqrt{10}\, [d^\dagger\otimes d]_{1k}
                   -\sqrt{5} \, [p^\dagger\otimes p]_{1k} \,, &\\
&\hat O_\nu = -\sqrt{10}\, [d^\dagger\otimes d]_{3\nu} 
                     +\sqrt{5} \, [p^\dagger\otimes p]_{3\nu}\,,&
\eea
where $[d^\dagger\otimes d]_{1k}$ signifies the SU(2)-coupled product
\be 
[d^\dagger\otimes d]_{1k}
 = \sum_{mn} (2, n, 2, m|1,k) d^\dag_m d_n \,.
\ee

These operators satisfy the commutation relations 
\bea 
{[}\hat L_k,\hat L_l{]}
 & = & \sqrt{2}\ (1 l,1 k|1 k+l)\, \hat L_{k+l}\, ,\\ 
{[}\hat L_k,\hat O_\nu{]}
 & = &  2\sqrt{3}\ (1 \nu,3 k |3 k+\nu)\,\hat O_{k+\nu}\, ,\\ 
{[}\hat O_\mu ,\hat O_\nu{]}
 & = & 2\sqrt{7}\ (3 \nu,3 \mu|1\mu+\nu)\,\hat L_{\mu+\nu}
         - \sqrt{6}\ (3 \nu,3 \mu |3\mu+\nu)\,\hat O_{\mu+\nu} \, .
\eea

Specification of the $\{d^\dag_M\}$ and $\{ p^\dag_M\}$ operators in terms
of the $\{\eta^\dag_\nu\}$ and $\{\zeta^\dag_m\}$ operators then defines
the embedding of SO(3) $\subset$ SO(5) as the subgroup
with Lie algebra $\so(3)$ spanned by the $\{ \hat L_k\}$ angular
momentum operators. 
From now on,  all references to SO(3) or $\so(3)$ will mean 
this subgroup or its Lie algebra.

\subsection{Relationships between the two bases}

Many choices of relationship are possible.
However, the simple relationship defined by $d^\dag_M = \eta^\dag_M$ and
$p^\dag_M =\zeta^\dag_M$ is unsatisfactory because, for this choice, it is found that
$\hat L_0= 2\hat X_0 + \hat F_0$; this means that $\hat L_0$ lies in the
intrinsic $\u(2)$ subalgebra and the conditions for Theorem 1 are violated.
A  satisfactory relationship is given by setting
\bea 
d^\dag_M = e^{\frac{\pi}{4}(\hat S_+-\hat S_-)} \eta^\dag_M
e^{-\frac{\pi}{4}(\hat S_+-\hat S_-)} ,\\
p^\dag_M = e^{\frac{\pi}{4}(\hat S_+-\hat S_-)} \zeta^\dag_M
e^{-\frac{\pi}{4}(\hat S_+-\hat S_-)}.
\eea
This relationship gives
\bea 
\eta^\dag_2
 = \textstyle\sqrt{\frac{1}{2}}\,\big(
d^\dag_2+d^\dag_{-1}\big), && 
\eta^\dag_{-2}
 = \textstyle\sqrt{\frac{1}{2}}\,\big(
d^\dag_{-2}-d^\dag_{1}\big), \\ 
&\eta^\dag_0
 = d^\dag_0 ,&\\
\eta^\dag_1
 = \textstyle\sqrt{\frac{1}{2}}\big( d^\dag_1+d^\dag_{-2}\big), && 
\eta^\dag_{-1}
 = \sqrt{\frac{1}{2}}\big(
d^\dag_{-1}-d^\dag_{2}\big),
\eea 
and
\bea 
\zeta^\dag_{\frac{3}{2}}
 = \textstyle\sqrt{\frac{1}{2}}\,\big(
p^\dag_{\frac{3}{2}}- p^\dag_{-\frac{3}{2}} \big), &\quad&
\zeta^\dag_{-\frac{3}{2}}
 =\textstyle \sqrt{\frac{1}{2}}\,\big(
p^\dag_{-\frac{3}{2}}+ p^\dag_{\frac{3}{2}} \big), \\
 \zeta^\dag_{\frac{1}{2}}
 = p^\dag_{\frac{1}{2}}, &\quad&
\zeta^\dag_{-\frac{1}{2}}
 = p^\dag_{-\frac{1}{2}}.
\eea
The relationship between the two bases for $\so(5)$ is then given by 
\bea 
&& \hat L_0 = \hf(\hat X_0-\hat F_0) - \textstyle\frac{3}{2}(\hat
S_++\hat S_-) ,\label{eq:55}\\ 
&& \hat L_\pm = 2\hat T_\pm + 
\textstyle\sqrt\frac{3}{2}\,(\hat F_\pm+\hat X_\mp) ,\label{eq:56}\\ 
&& \hat O_0 = \textstyle\frac{3}{2}(\hat X_0-\hat F_0)
 + \textstyle\frac{1}{2}(\hat S_++\hat S_-),\label{eq:57}\\ 
&& \hat O_{\pm 1} = \mp\sqrt{3}\,\hat T_\pm 
\pm \textstyle\sqrt\frac{1}{2}\,(\hat F_\pm +\hat X_\mp ) ,\\ 
&& \hat O_{\pm 2} = \textstyle\frac{\sqrt{5}}{2}\,(\hat
X_\pm -\hat F_\mp) ,\\
 && \hat O_{\pm 3} = \textstyle\frac{\sqrt{5}}{2}\, 
(\mp \hat X_0 \mp \hat F_0-\hat S_+ +\hat S_-) , \label{eq:60}
\eea
where $\hat L_{\pm}=\mp \sqrt{2}\hat L_{\pm 1}$.

\section{Construction of VCS basis wave functions} \label{sect:basis}

Orthonormal basis states $\{ |(vf)\tau LM\rangle\}$ for the (10) and
$(0\hf)$ irreps, for which the multiplicity index $\tau$ is redundant,
are given by
\bea 
&&|(10)2M\rangle = d^\dag_M |0\rangle,\quad M = 0,\pm 1, \pm 2 , \\
&&|(0\hf)\textstyle\frac{3}{2} M\rangle = p^\dag_M |0\rangle,\quad M = \pm \hf, \pm
\textstyle\frac{3}{2} . 
\eea 
Thus, from the definitions (\ref{eq:15})-(\ref{eq:18})), the corresponding VCS wave functions are given by
\bea 
\Psi^{(10)}_{2M}(\Omega) 
 &=&\xi^{(10)}_0 \langle 0|\eta^2 \hat R(\Omega)d^\dag_M |0\rangle
 = \sum_K\xi^{(10)}_0 \langle 0|\eta^2
d^\dag_K |0\rangle \mathcal{D}^2_{KM}(\Omega) \nonumber\\
 &=& \textstyle\frac{1}{\sqrt{2}} \xi^{(10)}_0 
\left[\mathcal{D}^2_{2M}(\Omega) + \mathcal{D}^2_{-1,M}(\Omega)\right] \,,
\eea
and
\bea\label{eq:fund4Dwf} 
\Psi^{(0 \frac{1}{2})}_{\frac{3}{2}M}(\Omega)
 &=& \xi^{(0\frac{1}{2})}_\frac{1}{2}
\langle 0|\zeta^\frac{3}{2}\hat R(\Omega) p^\dag_M |0\rangle
+\xi^{(0\frac{1}{2})}_{-\frac{1}{2}}
\langle 0|\zeta^\frac{1}{2}\hat R(\Omega) p^\dag_M |0\rangle
\nonumber\\ 
 &=& \textstyle\frac{1}{\sqrt{2}} \xi^{(0 \frac{1}{2})}_\frac{1}{2} 
\left[\mathcal{D}^\frac{3}{2}_{\frac{3}{2}M}(\Omega) -
\mathcal{D}^\frac{3}{2}_{-\frac{3}{2},M}(\Omega)\right]
+\xi^{(0 \frac{1}{2})}_{-\frac{1}{2}} 
\mathcal{D}^\frac{3}{2}_{\frac{1}{2}M}(\Omega) \,.
\eea
From Eq.\ (\ref{eq:15}), it follows that
\begin{equation}
  \xi^{(v_1f_1)}_{m_1}\odot \xi^{(v_2f_2)}_{m_2} = 
\sqrt{{(f+m)!\, (f-m)!\, v!\over (f_1+m_1)!\, (f_1-m_1)!\, (f_2+m_2)!\,
(f_2-m_2)!\, v_1!\, v_2!} }\ \xi^{(vf)}_m \,,
\end{equation}
where
\begin{equation}
 v=v_1+v_2\,, \quad f = f_1+f_2 \,, \quad m = m_1+m_2  \,.
\end{equation}
From the properties of the Wigner rotation matrices, it also follows
that
\begin{equation}
  \mathcal{D}^{L_1}_{K_1M_1} \odot \mathcal{D}^{L_2}_{K_2M_2} =
\sum_{L=|L_1-L_2|}^{L_1+L_2} (L_2M_2,L_1M_1 |LM) (L_2K_2,L_1K_1 |LK) \,
\mathcal{D}^L_{KM}
\,,
\end{equation}
where
\begin{equation}
  K= K_1+K_2 \,,\quad M=M_1+M_2 \,.
\end{equation}
Thus, Theorem 2 and these expressions lead naturally to an
algorithm for constructing (non-orthonormal) basis wave functions for an
arbitrary $\so(5)$ irrep.

It is convenient to start by constructing a non-orthonormal basis of VCS wave functions of the form
\begin{equation}
\Phi^{(vf)}_{\tau LM}= \sum_{mk} \xi^{(vf)}_m\, 
 b^{(vf)}_{mK} (\tau L)\, \mathcal{D}^L_{KM} \,, \label{eq:NonOrthBasis}
\end{equation}
with $ b^{(vf)}_{mK} (\tau L)$ coefficients conveniently chosen to be
real and normalised such that
\be 
\sum_{mK} b^{(vf)}_{mK} (\tau L)\,b^{(vf)}_{mK} (\sigma L) =
\delta_{\tau\sigma}. \label{eq:70}
\ee
The functions $\Phi$ so normalised will be related to the orthonormal $\so(5)$ functions $\Psi$ in 
section \ref{sec:trans}.

In order to carry out this basis construction efficiently, it is useful to know the
values of the angular momentum and the multiplicity of their occurrence in
any given irrep.
In other words, we need to know the $\so(5)\downarrow\so(3)$ branching rules which
give the $\so(3)$ irreps contained in any given $\so(5)$ irrep.
The $\so(5)\downarrow\so(3)$ branching rules for irreps of the type $(v0)$ were 
conveniently summarised by Williams and Pursey \cite{14} in the form
\begin{eqnarray} 
L&=& 2K,\; 2K-2,\; 2K-3,\; \ldots , \; K \nonumber\\
     K &=& v, \; v-3,\; v-6,\; \ldots ,\; K_{\rm min} \, ,
\end{eqnarray}
where $K_{\rm min} = 0$, 1, or 2.
The branching rules for the ireps of type $(0f)$ are  given \cite{Corr}
by
\begin{eqnarray} 
L&=& 3K,\; 3K-2,\; 3K-3,\; \ldots , \; K \nonumber\\
     K &=& f, \; f-2,\; f-4,\; \ldots ,\; K_{\rm min} \, ,
\end{eqnarray}
where $K_{\rm min} = 0$ or 1. 
The branching rules for a generic irrep can be inferred by use of
character theory \cite{Santo} or by a simple `peeling-off' program
\cite{Bahri} which uses knowledge of the number of eigenvalues of the
$\hat L_0$ operator.
The $\so(3)$ content of some low-dimensional $\so(5)$ irreps is given in
Table \ref{tab:so3}.
\begin{table}[!htb]
\caption{The $\so(3)$ content of some low-dimensional $\so(5)$ irreps;
note that the spinor irreps are labelled in the table by $2L$ for
convenience. \label{tab:so3}}
\begin{tabular}{l|l|l|l} 
\multicolumn{2}{c}{Genuine irreps} & \multicolumn{2}{c}{Spinor irreps} \\
\hline
$(v,f)$   & $L$                             &  $(v,f)$  & $2L$ \\
\hline
\hline
$(1,0)$   & $2$                             & $(0,1/2)$ & $3$ \\
$(0,1)$   & $1,3$                           & $(1,1/2)$ & $1,5,7$ \\
$(2,0)$   & $2,4$                           & $(0,3/2)$ & $3,5,9$ \\
$(1,1)$   & $1,2,3,4,5$                     & $(1,3/2)$ & $1,3,5,7,7,9,11,13$ \\
$(0,2)$   & $0,2,3,4,6$                     & $(2,1/2)$ & $3,5,7,9,11$ \\
$(3,0)$   & $0,3,4,6$                       & $(0,5/2)$ & $3,5,7,9,11,15$ \\
$(2,1)$   & $1,2,3,3,4,5,5,6,7$             & $(1,5/2)$ & $1,3,5,5,7,7,9,9,11,11,13,13,15,17,19$ \\
$(1,2)$   & $1,2,2,3,4,4,5,5,6,7,8$         & $(2,3/2)$ & $1,3,5,5,7,7,9,9,11,11,13,13,15,17$ \\
$(0,3)$   & $1,3,3,4,5,6,7,9$               & $(3,1/2)$ & $3,5,7,9,9,11,13,15$ \\
$(4,0)$   & $2,4,5,6,8$                     & $(0,7/2)$ & $3,5,7,9,9,11,13,15,17,21$ \\
\hline
\end{tabular}
\end{table}

A basis for any irrep of highest weight $(vf)$ can now be built up by
taking coupled products of the above basic wave functions.
For example, for the (20) irrep,
\be 
\Phi^{(20)}_{LM} \propto \left[ \Psi^{(10)}_2 \odot
\Psi^{(10)}_2\right]_{LM},\quad L=2,4 \,,
\ee
gives
\bea 
\Phi^{(20)}_{LM}
 & \propto & 
\xi^{(20)}_0 \sum_{K_1K_2} b^{(10)}_{0K_1} (2) b^{(10)}_{0K_2} (2)\, (2K_1,2K_2 |L,K_1+K_2) 
\,\mathcal{D}^L_{K_1+K_2,M} \nonumber \\
 & \propto & 
\xi^{(20)}_0 \left[(22,22|L4)\,\mathcal{D}^L_{4M} + 2(22,2,-1|L1)\,\mathcal{D}^L_{1M} + 
(2,-1,2,-1|L,-2)\,\mathcal{D}^L_{-2,M} \right] .
\eea
Thus, we obtain the results shown in Table \ref{tab:1}.
\begin{table}
\caption{The $b^{(v0)}_{0K}(\tau L)$ coefficients for some irreps of type
$(v0)$.   Note that here and in the following tables the coefficients are 
normalised according to Eq. (\ref{eq:70}) and their phases are chosen so that  
the leading nonzero coefficient for each $(L,\tau)$ is
positive.  \label{tab:1}}
$$\begin {array}{c|cc} 
\multicolumn{3}{c} {$$(vf)=(10)$$}\\
\hline\hline
 & \multicolumn{2}{c} {K} \\
L & 2 & -1 \\ \hline
  2 & 1/\sqrt{2}&1/\sqrt{2} \\
\hline\hline
\end {array}$$
$$\begin {array}{c|ccc} 
\multicolumn{4}{c} {$$(vf)=(20)$$}\\
\hline\hline
 & \multicolumn{3}{c} {K} \\
L & 4 & 1 & -2 \\ \hline
  2 & 0& 2/\sqrt{5}&-1/\sqrt{5} \\
4 &\sqrt{7/13} &\sqrt{2/13}&2/\sqrt{13} \\
\hline\hline
\end {array}$$
$$\begin {array}{c|cccc} 
\multicolumn{5}{c} {$$(vf)=(30)$$}\\
\hline\hline
 & \multicolumn{4}{c} {K} \\
L & 6 & 3 & 0&-3 \\ \hline
  0 &  0&0&1&  0\\
3&0 &\sqrt{2/5} & -\sqrt{1/2}&-\sqrt{1/10} \\
4 &0&2 \sqrt{5/32} &\sqrt{7/32}& -\sqrt{5/32} \\
6& \sqrt{77/124} & 3\sqrt{7/620} & \sqrt{3/31}& 2\sqrt{7/155}   \\
\hline\hline
\end {array}$$
\end{table}
Similarly, we have for the $(01)$ irrep
\bea 
\Phi^{(01)}_{LM} & \propto & \xi^{(01)}_1 \sum_{K_1K_2}
b^{(0\frac{1}{2})}_{\frac{1}{2}K_1}(\textstyle\frac{3}{2})\,
b^{(0\frac{1}{2})}_{\frac{1}{2}K_2}(\textstyle\frac{3}{2}) \,
\textstyle(\frac{3}{2} K_1,\frac{3}{2} K_2|L,K_1+K_2) 
\,\mathcal{D}^L_{K_1+K_2,M}\nonumber\\
&&+\sqrt{2}\,\xi^{(01)}_0 \sum_{K_1K_2}
b^{(0\frac{1}{2})}_{-\frac{1}{2}K_1}(\textstyle\frac{3}{2})\,
b^{(0\frac{1}{2})}_{\frac{1}{2}K_2}(\textstyle\frac{3}{2}) \,
\textstyle(\frac{3}{2} K_1,\frac{3}{2} K_2|L,K_1+K_2) 
\,\mathcal{D}^L_{K_1+K_2,M} \nonumber\\
&&+\xi^{(01)}_{-1} \sum_{K_1K_2}
b^{(0\frac{1}{2})}_{-\frac{1}{2}K_1}(\textstyle\frac{3}{2})\,
b^{(0\frac{1}{2})}_{-\frac{1}{2}K_2}(\textstyle\frac{3}{2}) \,
\textstyle(\frac{3}{2} K_1,\frac{3}{2} K_2|L,K_1+K_2) 
\,\mathcal{D}^L_{K_1+K_2,M} \,,
\eea
which leads to a set of coefficients given in Table \ref{tab:2}. 

\begin{table}
\caption{The $b^{(0f)}_{mK}(\tau L)$ coefficients for some irreps of type
$(0f)$.  Note that for an $\so(5)$ irrep $(vf)$ the coefficients among 
different $\so(3)$ irreps can all be indexed by the same values of $K-m$, hence 
this is used to label columns. \label{tab:2}}
$$\begin {array}{cc|cc} 
\multicolumn{4}{c} {$$(vf)=(0\hf)$$}\\ 
\hline\hline
 & & \multicolumn{2}{c} {K-m} \\
L& m & 1 & -2 \\ \hline
\textstyle\frac{3}{2}& +\frac{1}{2} & 1/2 & -1/2  \\ 
&-\frac{1}{2}  &1/\sqrt{2}  \\
\hline\hline
\end {array}$$
$$\begin {array}{cc|ccc} 
\multicolumn{5}{c} {$$(vf)=(01)$$}\\
\hline\hline
 & & \multicolumn{3}{c} {K-m} \\
L& m & 2 & -1 &-4\\ \hline
1& +1 & 0&3/\sqrt{23}&0  \\
& \;\;\; 0  &0&\sqrt{6/23} \\
&-1&2\sqrt{2/23}\\ 
\hline
3& +1 & \sqrt{5/37}&-1/\sqrt{37}&\sqrt{5/37}  \\
& \;\;\; 0  &\sqrt{10/37}&-2/\sqrt{37}  \\
&-1&2\sqrt{3/37}\\
\hline\hline
\end {array}$$
$$\begin {array}{cc|cccc} 
\multicolumn{6}{c} {$$(vf)=(0\textstyle\frac{3}{2})$$}\\
\hline\hline
 & & \multicolumn{4}{c} {K-m} \\
L& m & 3 & 0&-3&-6 \\ \hline
\noalign{\smallskip}
\textstyle\frac{3}{2}& +\frac{3}{2} & 0 & \sqrt{27/152}&\sqrt{27/152} &0  \\
& +\frac{1}{2}  & 0 & 3\sqrt{2/152}&0   \\
& -\frac{1}{2}  & 0 & 4\sqrt{3/152}  \\
\textstyle & -\frac{3}{2}  & -4\sqrt{2/152}   \\ 
\noalign{\smallskip}
\hline 
\noalign{\smallskip}\textstyle
\frac{5}{2}& +\frac{3}{2} & 0 & \sqrt{27/232}&-\sqrt{27/232} &0  \\
& +\frac{1}{2}  & 0 & 4\sqrt{3/232}&-\sqrt{30/232}   \\
& -\frac{1}{2}  & 2\sqrt{5/232} & 2\sqrt{2/232}  \\
\textstyle & -\frac{3}{2}  & 6\sqrt{2/232}   \\ 
\noalign{\smallskip}
\hline
\noalign{\smallskip}
\frac{9}{2}& +\frac{3}{2} & 2\sqrt{7/328} & -\sqrt{3/328}&\sqrt{3/328} &
-2\sqrt{7/328}  \\ & +\frac{1}{2}  & 2\sqrt{14/328} & -4\sqrt{1/328}&
\sqrt{14/328}   \\ & -\frac{1}{2}  & 2\sqrt{21/328} & -2\sqrt{6/328}  \\
\textstyle & -\frac{3}{2}  & 6\sqrt{2/328} \\
\noalign{\smallskip}
\hline\hline
\end {array}$$
\end{table}

In general we can form basis states for an irrep $(vf)$ from the coupled
products
\bea
\left[\Phi^{(v0)}_{\tau_2 L_2} \odot \Phi^{(0f)}_{\tau_1 L_1}\right]_{LM}
 = \sum_m \xi^{(v0)}_0\odot \xi^{(f0)}_m
\sum_{K_1K_2} b^{(v0)}_{0K_2} (\tau_2 L_2)\, b^{(0f)}_{mK_1} (\tau_1 L_1)\, 
(L_1K_1,L_2K_2|L,K_1+K+2) \,\mathcal{D}^L_{K_1+K_2,M}.
\eea
Some examples are given in Table \ref{tab:3}.
\begin{table}  
\caption{The $b^{(vf)}_{mK}(\tau L)$ coefficients for some
generic irreps. \label{tab:3}}
$$\begin {array}{cc|ccc} 
\multicolumn{5}{c} {$$(vf)=(1\hf)$$}\\
\hline\hline 
 & & \multicolumn{3}{c} {K-m} \\
L & m & 3 & 0 & -3 \\ 
\hline
\hf & +\hf & 0 & \sqrt{3/5} & 0 \\
    & -\hf& 0 &-\sqrt{2/5} &   \\
\hline
5\over 2 &+\hf & 0 & \sqrt{15/47} & \sqrt{6/47} \\
         &-\hf & -4/\sqrt{47} & \sqrt{10/47} &  \\
\hline
7\over 2 &+\hf & \sqrt{7/22} & 1/\sqrt{110} & -\sqrt{2/11} \\
         &-\hf & \sqrt{2/11} & \sqrt{6/55} &  \\
\hline\hline
\end {array}$$
$$\begin{array}{cc|cccc}
\multicolumn{6}{c}{$$(vf)=(11)$$}\\
\hline\hline
 & & \multicolumn{4}{c} {K-m} \\
L & m & 4 & 1 & -2 & -5 \\
\hline
1 & +1 & 0 & 0 & 3/\sqrt{29} & 0 \\
  &\phantom{+}0 & 0 & -2\sqrt{3/29} & 0 & \\
  & -1 & 0 & -2\sqrt{2/29} & & \\
\hline
2 & +1 & 0 & 2\sqrt{3/31} & -\sqrt{3/31} & 0 \\
  &\phantom{+}0 & 0 & 2/\sqrt{31} & 2/\sqrt{31} & \\
  & -1 & 0 & -2\sqrt{2/31} & & \\
\hline
3 & +1 & 0 & \sqrt{15/109} & 2\sqrt{6/109} & 0 \\
  &\phantom{+}0 & 0 & \sqrt{2/109} & 2\sqrt{5/109} & \\
  & -1 & 2\sqrt{10/109} & 2\sqrt{2/109} & & \\
\hline
4 & +1 & 0 & 5\sqrt{3/829} & -2\sqrt{6/829} & -2\sqrt{21/829} \\
  &\phantom{+}0 & -4\sqrt{7/829} & 11\sqrt{2/829} & -2/\sqrt{829} & \\
  & -1 & -2\sqrt{42/829} & 2\sqrt{30/829} & & \\
\hline
5 & +1 & \sqrt{70/313} & 0 & -\sqrt{3/313} & 2\sqrt{7/313} \\
  &\phantom{+}0 & 2\sqrt{21/313} & 2/\sqrt{313} & -2\sqrt{7/313} & \\
  & -1 & 2\sqrt{14/313} & 2\sqrt{10/313} & & \\
\hline\hline
\end{array}$$ 
\end{table}

It is important to recognise that the basis $\{ \Phi^{(vf)}_{\tau LM}\}$
is not orthornormal relative to the appropriate inner product for VCS wave
functions. An orthonormal basis is one relative to which the VCS
representations are unitary; i.e., for which the elements of the $\so(5)$
Lie algebra are represented by Hermitian operators.

\section{VCS representation of the $\so(5)$ algebra} \label{sect:irreps}

According to Eq.\ (\ref{eq:10}), the action of the angular
momentum operators on any SO(3)-coupled VCS wave functions of the form
\be 
\Phi^{(vf)}_{\tau LM}(\Omega) = \sum_{mK} \xi^{(vf)}_m \langle (vf)m|
(vf)\tau LK\rangle \, \mathcal{D}^L_{KM}(\Omega)
 = \sum_{mK}\xi^{(vf)}_m b^{(vf)}_{mK}(\tau L) \,
\mathcal{D}^L_{KM}(\Omega) \label{eq:77}
\ee
 is given by the standard action
\bea 
[\Gamma(L_0) \Phi^{(vf)} _{\tau LM} ] (\Omega) &=& M\Phi^{(vf)}
_{\tau LM}  (\Omega) \,, \label{eq:78}\\
{[}\Gamma(L_\pm) \Phi^{(vf)}_{\tau LM} ] (\Omega)
&=&\sqrt{(L\mp M) L\pm M+1)}\,\Phi^{(vf)}_{\tau L,M\pm 1} 
(\Omega)\,. \label{eq:79}
\eea
The action of the octupole operators is given by
\be 
[\Gamma(O_\nu) \Phi^{(vf)} _{\tau LM} ] (\Omega) =
\sum_{mK\mu} \xi^{(vf)}_m \langle (vf)m|\hat O_\mu
|(vf)\tau LK\rangle \, 
\mathcal{D}^3_{\mu\nu}(\Omega)\mathcal{D}^L_{KM}(\Omega) ,
\ee
or, in coupled form, 
\be 
\left[\Gamma(O) \otimes\Phi^{(vf)}_{\tau L} \right]_{L'M}=
\sum_{mK\mu}
\xi^{(vf)}_m \langle (vf)m|\hat O_\mu |(vf)\tau LK\rangle \, (LK,3\mu
|L',K+\mu)\,
\mathcal{D}^{L'}_{K+\mu,M}. \label{eq:81}
\ee
Thus, it remains to determine the matrix elements
$\{ \langle (vf)m|\hat O_\mu|(vf)\tau LK\rangle\}$ to define an $\so(5)$ irrep 
$(vf)$. From the definition of the states $\{|(vf)m\rangle\}$ as highest grade
states, cf.\ Eqns.\ (\ref{eq:1})-(\ref{eq:3}), we have the identities
\bea
 & \langle (vf)m|\hat S_-|(vf)\tau LK\rangle =
\langle (vf)m|\hat X_-|(vf)\tau LK\rangle =
\langle (vf)m|\hat T_-|(vf)\tau LK\rangle = 0, \\
&\langle (vf)m|\hat X_0|(vf)\tau LK\rangle = (v+f)\langle (vf)m|(vf)\tau
LK\rangle , \quad  
\langle (vf)m|\hat F_0|(vf)\tau LK\rangle = m\langle
vfm|(vf)\tau LK\rangle ,  & \\
&\langle (vf)m|\hat F_\pm|(vf)\tau LK\rangle = 
\sqrt{(f\pm m)( f\mp m+1)}\ \langle (vf)m\mp 1|(vf)\tau LK\rangle .&
\eea
and from the standard action of the angular momentum operators on the
states $\{ |(vf)\tau LK\rangle\}$, we have
\bea
 &\langle (vf)m|\hat L_0|(vf)\tau LK\rangle = 
K\langle (vf)m|(vf)\tau LK\rangle ,&\\
&\langle (vf)m|\hat L_\pm|(vf)\tau LK\rangle = 
\sqrt{(L\mp K)(L\pm K+1)}\ \langle (vf)m|(vf)\tau L,K\pm 1\rangle .&
\eea
Eqs.\ (\ref{eq:55}-\ref{eq:56}) give the relationships 
\bea
 &&\hat S_+ =\frac{1}{3}(\hat X_0-\hat F_0) - \hat S_- -
\frac{2}{3}\hat L_0 ,\\
&&\hat X_+ = \sqrt\frac{2}{3} \hat L_- -2\sqrt\frac{2}{3}\hat T_- - \hat
F_- ,\\
&&\hat T_+ = \frac{1}{2} \hat L_+ -\frac{1}{2}\sqrt\frac{3}{2}(\hat F_+ +
\hat X_-)
\eea
which makes it possible to rewrite Eqs.\ (\ref{eq:57}-\ref{eq:60}) in the
form
\bea
 &&\hat O_0 = \frac{5}{3} (\hat X_0-\hat F_0)- \frac{1}{3}\hat L_0 ,\\
&&\hat O_1= -\frac{\sqrt{3}}{2} \hat L_+ + \frac{5}{2\sqrt{2}} (\hat F_+ 
+ \hat X_-) ,\\
&&\hat O_{-1}= -\frac{1}{\sqrt{3}} \hat L_- + \frac{5}{\sqrt{3}} \hat
T_-  ,\\
&&\hat O_2= \sqrt\frac{5}{6} \hat L_- -\sqrt\frac{10}{3} \hat T_-
-\sqrt{5}\, \hat F_-,\\
&&O_{-2}=\frac{\sqrt{5}}{2}\, \hat X_- -\frac{\sqrt{5}}{2}\, \hat F_+,\\
&&O_3= -\frac{2\sqrt{5}}{3} \hat X_0 -\frac{\sqrt{5}}{3} \hat F_0
+ \sqrt{5}\, \hat S_- + \frac{\sqrt{5}}{3}\, \hat L_0 ,\\
&&O_{-3}=\frac{\sqrt{5}}{3} \hat X_0+\frac{2\sqrt{5}}{3} \hat F_0
+ \sqrt{5}\, \hat S_- + \frac{\sqrt{5}}{3}\, \hat L_0 .
\eea
We now have all of the $\so(5)$ operators in terms of the basis operators
$\{ S_-,T_-,X_-,X_0,F_0,F_\pm,L_0,L_\pm \}$ with known
algebraic actions on either the highest grade states or on the $\so(3)$-coupled states.  
Note that Theorem 1 ensures that this set is linearly independent and spans $\so(5)$.  
This is a general feature of a useful VCS construction.

We obtain
\bea  
\langle (vf)m|\hat O_0|(vf)\tau LK\rangle &=&
\frac{1}{3} (5v+5f-5m-K)\, b^{(vf)}_{mK}(\tau L) ,
\\ 
\langle (vf)m|\hat O_1|(vf)\tau LK\rangle &=&
\textstyle \frac{5}{2} \sqrt{\hf(f+m)(f-m+1)}\  
b^{(vf)}_{m-1,K}(\tau L)   \nonumber\\ 
&&\textstyle  -\frac{1}{2} \sqrt{3(L-K)(L+K+1)}\ 
 b^{(vf)}_{m,K+1}(\tau L)  ,
\\
\langle (vf)m|\hat O_{-1}|(vf)\tau LK\rangle  &=&
\textstyle  - \sqrt{\frac{1}{3}(L+K)(L-K+1)}\ 
 b^{(vf)}_{m,K-1}(\tau L)   ,
\\
\langle (vf)m|\hat O_{2}|(vf)\tau LK\rangle &=&
\textstyle \sqrt{\frac{5}{6}(L+K)(L-K+1)}\
b^{(vf)}_{m,K-1}(\tau L)  \nonumber\\
&&  -\sqrt{5(f-m)(f+m+1)}\
 b^{(vf)}_{m+1,K}(\tau L),
\\ 
\langle (vf)m|\hat O_{-2}|(vf)\tau LK\rangle &=&
\textstyle  - \frac{1}{2}\, 
\sqrt{5(f+m)(f-m+1)}\  b^{(vf)}_{m-1,K}(\tau L),
\\ 
\langle (vf)m|\hat O_{3}|(vf)\tau LK\rangle  &=&
\textstyle -\frac{\sqrt{5}}{3} (2v+2f +m-K)\, b^{(vf)}_{mK}(\tau L) ,
\\ 
\langle (vf)m|\hat O_{-3}|(vf)\tau LK\rangle  &=&
\textstyle \frac{\sqrt{5}}{3} (v+f+2m+K)\, b^{(vf)}_{mK}(\tau L) .
\eea
It follows that Eq.\ (\ref{eq:81}) has the explicit expansion given by
\be \left[\Gamma(O) \otimes\Phi^{(vf)}_{\tau L} \right]_{L'M}=
\sum_{mKm'K'} b^{(vf)}_{mK}(\tau L)\, M^{(vf)}_{mKL,m'K'L'}\,
\xi^{(vf)}_{m'} 
\mathcal{D}^{L'}_{K'M}, \label{eq:104}
\ee
where $M^{(vf)}$ is a matrix with non-zero entries
\bea\label{eq:Mmatrix}  
M^{(vf)}_{mKL,mKL'} &=& 
\textstyle\frac{1}{3} (5v+5f-5m-K) (LK,3\;0|L'K) \nonumber\\
&& -\hf\sqrt{3(L+K)(L-K+1)}\ (L\;K-1,3\;1|L'K)\nonumber\\
&&\textstyle - \sqrt{\frac{1}{3}(L-K)(L+K+1)}\
(L\;K+1,3\;-1|L'K),\nonumber\\ 
M^{(vf)}_{mKL,m+1\;K+1\;L'} &=&
\textstyle \frac{5}{2} \sqrt{\hf(f-m)(f+m+1)}\ 
(LK,3\;1|L'\;K+1),\nonumber\\ 
M^{(vf)}_{mKL,m-1\;K+2\;L'} &=&
\textstyle - \sqrt{5(f+m)(f-m+1)}\  (LK,3\;2|L'\;K+2),\nonumber\\
M^{(vf)}_{mKL,m\;K+3\;L'} &=&
\textstyle \sqrt{\frac{5}{6}(L-K)(L+K+1)}\ 
(L\;K+1,3\;2|L'\;K+3)\nonumber\\ 
&& - \frac{\sqrt{5}}{3}
(2v+2f+m-K)\,(LK,3\;3|L'\;K+3),\nonumber\\ 
M^{(vf)}_{mKL,m+1\;K-2\;L'} &=&
\textstyle - \hf\sqrt{5(f-m)(f+m+1)}\  (LK,3\;-2|L'\;K-2),\nonumber\\
M^{(vf)}_{mKL,m\;K-3\;L'} &=&
\textstyle \frac{\sqrt{5}}{3} (v+f+2m+K)\,(LK,3\;-3|L'\;K-3) \,. \label{eq:108}
\eea
Now, if the basis wave functions $\{ \Phi^{(vf)}_{\tau LM}\}$ are chosen,
as defined in Sect.\ \ref{sect:basis}, with $\{b^{(vf)}_{mK}(\tau L)\}$
coefficients that are real and satisfy Eq.\ (\ref{eq:70}), it follows that
Eq.\ (\ref{eq:104}) can be expressed
\be 
\left[\Gamma(O) \otimes\Phi^{(vf)}_{\tau L} \right]_{L'M}=
\sum_{\sigma} \Phi^{(vf)}_{\sigma L'M} \mathcal{O}^{(vf)}_{\sigma L',\tau
L} ,
\ee
where
\be  
\mathcal{O}^{(vf)}_{\sigma L',\tau L} = \sum_{mKm'K'} 
b^{(vf)}_{mK}(\sigma L')\, M^{(vf)}_{mKL,m'K'L'}\,
b^{(vf)}_{m'K'}(\tau L) .\label{eq:107}
\ee
Thus, together with Eqs.\ (\ref{eq:78}-\ref{eq:79}), these equations give
the explicit transformations  of the 
 $\{ \Phi^{(vf)}_{\tau LM}\}$ basis wave functions for any $\so(5)$
irrep.

Unfortunately, these matrices do not satisfy the Hermiticity conditions
required of a unitary representation.
This is the because the $\{ \Phi^{(vf)}_{\tau LM}\}$ basis is not
orthonormal relative to the appropriate VCS inner product.

\section{The matrices of unitary irreps}\label{sec:trans} 

We now suppose that  $\{ |(vf)\alpha LM\rangle \}$ is an orthonormal
basis for an irrep $(vf)$ and that the corresponding VCS wave functions 
have expansions
\be 
\Psi^{(vf)}_{\alpha LM} = \sum_{mK}
 \xi^{(vf)}_m \langle (vf)m|(vf)\alpha LK\rangle \,
\mathcal{D}^L_{KM} = \sum_{mK}\xi^{(vf)}_m a^{(vf)}_{mK}(\alpha L) 
\,\mathcal{D}^L_{KM} .
\ee
Note that for an orthonormal basis we have called the expansion
coefficients of the VCS wave functions $a^{(vf)}_{mK}$ in order to distinguish them from the 
$b^{(vf)}_{mK}$ coefficients of the non-orthonormal $\Phi^{(vf)}_{\sigma LM}$ basis. The
Wigner-Eckart theorem for matrix elements in an orthonormal basis
\be 
\langle (vf)\beta L'M' |\hat O_\nu | (vf)\alpha LM\rangle =
\frac{1}{\sqrt{2L'+1}}\, (LM,3\nu |L'M')\, \langle (vf)\beta L'\| \hat O
\|(vf)\alpha L\rangle
\ee
then implies that
\be
\big[\Gamma(O) \otimes \Psi^{(vf)}_{ \alpha L} \big]_{L'M'} =
\frac{1}{\sqrt{2L'+1}}\,\sum_\beta \Psi^{(vf)}_{ \beta L'}\,  
\langle (vf) \beta L'\| \hat O \|(vf) \alpha L\rangle  \,, 
\ee
It follows from Eq.\ (\ref{eq:107}) that if the orthornormal basis wave
functions are expanded
\begin{equation}\label{eq:psiphi}
\Psi^{(vf)}_{ \alpha LM} = \sum_{ \sigma}
\Phi^{(vf)}_{ \sigma LM}\, K^{(vf)}_{ \sigma\alpha}(L) 
\end{equation}
then the desired  reduced matrix elements, relative to the orthonormal
basis, are given by
\be 
\langle (vf) \beta L'\| \hat O \| (vf) \alpha L\rangle
 = \sqrt{2L'+1}\, \sum_{ \sigma\tau} \bar K^{(vf)}_{ \beta\tau}\,
\mathcal{O}^{(vf)}_{ \tau L', \sigma L}\, K^{(vf)}_{\sigma\alpha} ,
\ee
where ${\bar K}^(vf)$ is the inverse of the  matrix $K^{(vf)}$, i.e., it is
defined such that
\be 
\sum_\sigma \bar K^{(vf)}_{\beta\sigma}
K^{(vf)}_{\sigma\alpha} =
\delta_{\alpha\beta}.
\ee
The $K^{(vf)}(L)$ matrices are chosen in VCS theory \cite{VCS} such that the
reduced matrix elements satisfy the Hermiticity condition
\be 
\langle (vf) \alpha L'\| \hat O \|(vf) \beta L\rangle^* = (-1)^{L-L'}
\langle (vf) \beta L\| \hat O \|(vf) \alpha L'\rangle \label{eq:114}
\ee
required of a unitary representation.  Such a transformation is found in 
two steps \cite{Kmatrix}. 
 
First observe that the submatrices $\mathcal{O}(L)$ with elements
\be 
\mathcal{O}_{ \sigma\tau}(L) = \mathcal{O}^{(vf)}_{ \sigma L, \tau L}
\ee
are real and symmetric.
Thus, we first make an orthogonal transformation 
\begin{equation}\label{eq:diag}
\tilde\Phi^{(vf)}_{ \alpha LM} = \sum_{\sigma}
\Phi^{(vf)}_{\sigma LM}\, \mathcal{K}^{(vf)}_{\sigma  \alpha}(L) \,,
\end{equation}
which diagonalises the $\mathcal{O}(L)$ matrices. The transformed matrix 
\begin{equation}
\tilde\mathcal{O}^{(vf)}_{ \beta L', \alpha L}
 = \sum_{ \sigma \tau}
\mathcal{K}^{(vf)}_{ \tau \beta}(L') \,
\mathcal{O}^{(vf)}_{ \tau L', \sigma L}
\mathcal{K}^{(vf)}_{ \sigma \alpha}(L) \,,
\end{equation}
then satisfies the equality
\begin{equation}
 \tilde\mathcal{O}^{(vf)}_{ \beta L, \alpha L} = \delta_{ \alpha \beta} \,
 \mathcal{O}^{(vf)}_{ \beta L, \beta L} \,.
\end{equation}
Before proceeding, it is important to note that states of different $L$
and $M$ are automatically orthogonal by virtue of their transformation
properties under $SO(3)$.
Moreover, if the reduced matrices for $L=L'$ are diagonal, this subset of
matrices automatically satisfies the Hermiticity condition
(\ref{eq:114}).

Thus, in general, it only remains to apply suitable scale factors to the
basis vectors
$\{\tilde\Phi^{(vf)}_{ \alpha LM}\}$ to obtain an orthonormal basis.
The required scale factors $\{ k^{(vf)}_{ \alpha L}\}$ must be such that
\begin{equation}
\big( k^{(vf)}_{ \alpha L}\big)^{-1}\tilde\mathcal{O}^{(vf)}_{ \alpha L, \beta
L'} k^{(vf)}_{ \beta L'} = (-1)^{L-L'} 
\big( k^{(vf)}_{ \beta L'}\big)^{-1}\tilde\mathcal{O}^{(vf)}_{ \beta
L', \alpha L} k^{(vf)}_{ \alpha L}
\end{equation}
The desired $K^{(vf)}$ matrices are then given by
\begin{equation}
 K^{(vf)}_{\tau \alpha}(L) = 
\mathcal{K}^{(vf)}_{\tau \alpha}(L)\, k^{(vf)}_{ \alpha L}
\end{equation}
with
\begin{equation}
  \left| {k^{(vf)}_{ \alpha L}\over k^{(vf)}_{ \beta L'}}\right|^2 =
(-1)^{L-L'} {\tilde\mathcal{O}^{(vf)}_{ \alpha L, \beta L'}\over
\tilde\mathcal{O}^{(vf)}_{ \beta L', \alpha L}} \,.
\end{equation}

Special consideration must be given to the relatively few irreps for
which there is a multiplicity of states of $L=0$, $\half$ or $1$.
This is because the $\mathcal{O}(L)$ matrices are identically zero for these $L$
values.
However, to ensure the orthogonality and correct normalisation of, for example,
the $L=0$ states it is sufficient to determine linear combinations
of these states which satisfy the equations
\be \langle (vf)\alpha 0\| \hat O\|(vf) \beta L\rangle = (-1)^L 
\langle (vf)\beta L\| \hat O\|(vf) \alpha 0\rangle ,
\ee
where $\{ |(vf)\beta LM\rangle \}$ is a small subset of states that have already
been orthonormalised.

\section{Sample results}\label{sec:results}

In this section we tabulate the a-coefficients $a^{(vf)}_{mK}(\tau L)$ of the unitary 
basis wave functions and the unitary SO(3)-reduced $\so(5)$ matrix elements of the 
$\hat{O}$ operator for the simplest generic $\so(5)$ irreps as well as the table for 
the first irrep with a multiplicity, namely $(1 \frac{3}{2})$, (we index the multiple $\so(3)$ 
irreps as $L_\tau$).  Because of the SO(3) 
reduction, we need only provide values between states of different $L$, since the 
Wigner-Ekart theorem provides the rest. Note that the matrix elements of the 
$\hat{L}$ operator are easily computed using Eqs. (\ref{eq:78}) 
and (\ref{eq:79}), and so they are not included in the tables.  Since the diagonalisation 
of Eq. (\ref{eq:diag}) must be done numerically, the values are given in floating point 
form.

\begin{table}[htb]
\caption{ Expansion coefficients and reduced matrix elements for $\so(5)$
irrep $(1,\hf)$.}
\begin{small}
\begin{tabular}{c|ccc|c|ccc}
\multicolumn{4}{c}{a-coefficients} & \multicolumn{4}{c}{reduced $\hat{O}$ matrix elements} \\
 & \multicolumn{3}{c}{$K-m$} & & & & \\
$m$ & $3$ & $0$ & $-3$ & $L$ & $\frac{1}{2}$ & $\frac{5}{2}$ & $\frac{7}{2}$ \\
\hline \hline
$\frac{1}{2}$ & 0 & 0.474341 & 0 & 
\multirow{2}{*}{$\frac{1}{2}$} & 
\multirow{2}{*}{$0$} & 
\multirow{2}{*}{$6.123724$} & 
\multirow{2}{*}{$5.999999$} \\
-$\frac{1}{2}$ & 0 & -0.387298 & 0 & & & & \\
\hline
$\frac{1}{2}$ & 0 & 0.517549 & 0.327326 & 
\multirow{2}{*}{$\frac{5}{2}$} & 
\multirow{2}{*}{$6.123724$} & 
\multirow{2}{*}{$2.535462$} & 
\multirow{2}{*}{$11.338934$} \\
-$\frac{1}{2}$ & -0.534522 & 0.422577 & 0 & & & & \\
\hline
$\frac{1}{2}$ & 0.499999 & 0.084515 & -0.377964 & 
\multirow{2}{*}{$\frac{7}{2}$} & 
\multirow{2}{*}{$-5.999999$} & 
\multirow{2}{*}{$-11.338934$} & 
\multirow{2}{*}{$3.070597$} \\
-$\frac{1}{2}$ & 0.462910 & 0.414039 & 0 & & & & \\
\end{tabular}
\end{small}
\end{table}

\begin{table}[htb]
\caption{ Expansion coefficients and reduced matrix elements for $\so(5)$
irrep $(2,\half)$. }
\begin{footnotesize}
\begin{tabular}{c|cccc|c|ccccc}
\multicolumn{5}{c}{a-coefficients} & \multicolumn{6}{c}{reduced $\hat{O}$ matrix elements} \\
 & \multicolumn{4}{c}{$K-m$} & & & & & & \\
$m$ & $5$ & $2$ & $-1$ & $-4$ & $L$ & $\frac{3}{2}$ & $\frac{5}{2}$ & $\frac{7}{2}$ & $\frac{9}{2}$ & $\frac{11}{2}$ \\
\hline \hline
$\frac{1}{2}$ & 0 & 0 & 0.439155 & 0 & 
\multirow{2}{*}{$\frac{3}{2}$} & 
\multirow{2}{*}{$4.225771$} & 
\multirow{2}{*}{$9.486832$} & 
\multirow{2}{*}{$10.954451$} & 
\multirow{2}{*}{$-6.866065$} & 
\multirow{2}{*}{$0$} \\
-$\frac{1}{2}$ & 0 & 0.414039 & 0.207019 & 0 & & & & & & \\
\hline
$\frac{1}{2}$ & 0 & 0.365148 & -0.204124 & 0.241522 & 
\multirow{2}{*}{$\frac{5}{2}$} & 
\multirow{2}{*}{$-9.486832$} & 
\multirow{2}{*}{$11.409582$} & 
\multirow{2}{*}{$-1.380131$} & 
\multirow{2}{*}{$8.391463$} & 
\multirow{2}{*}{$9.486832$} \\
-$\frac{1}{2}$ & 0 & -0.158113 & -0.316227 & 0 & & & & & & \\
\hline
$\frac{1}{2}$ & 0 & 0.365148 & -0.204124 & 0.241522 & 
\multirow{2}{*}{$\frac{7}{2}$} & 
\multirow{2}{*}{$10.954451$} & 
\multirow{2}{*}{$1.380131$} & 
\multirow{2}{*}{$-11.258858$} & 
\multirow{2}{*}{$-8.164965$} & 
\multirow{2}{*}{$11.775681$} \\
-$\frac{1}{2}$ & 0 & 0.516397 & -0.129099 & 0 & & & & & & \\
\hline
$\frac{1}{2}$ & 0 & 0.217597 & 0.255883 & 0.174077 & 
\multirow{2}{*}{$\frac{9}{2}$} & 
\multirow{2}{*}{$6.866065$} & 
\multirow{2}{*}{$8.391463$} & 
\multirow{2}{*}{$8.164965$} & 
\multirow{2}{*}{$12.465753$} & 
\multirow{2}{*}{$11.742179$} \\
-$\frac{1}{2}$ & -0.426401 & 0.023262 & 0.232621 & 0 & & & & & & \\
\hline
$\frac{1}{2}$ & 0.353553 & 0.082572 & 0 & -0.190692 & 
\multirow{2}{*}{$\frac{11}{2}$} & 
\multirow{2}{*}{$0$} & 
\multirow{2}{*}{$-9.486832$} & 
\multirow{2}{*}{$11.775681$} & 
\multirow{2}{*}{$-11.742179$} & 
\multirow{2}{*}{$8.628704$} \\
-$\frac{1}{2}$ & 0.261116 & 0.190692 & 0.190692 & 0 & & & & & & \\
\end{tabular}
\end{footnotesize}
\end{table}

\begin{table}[htb]
\caption{ Expansion coefficients and reduced matrix elements for $\so(5)$
irrep $(1,1)$.}
\begin{footnotesize}
\begin{tabular}{c|cccc|c|ccccc}
\multicolumn{5}{c}{a-coefficients} & \multicolumn{6}{c}{reduced $\hat{O}$ matrix elements} \\
 & \multicolumn{4}{c}{$K-m$} & & & & & & \\
$m$ & $4$ & $1$ & $-2$ & $-5$ & $L$ & $1$ & $2$ & $3$ & $4$ & $5$ \\
\hline \hline
1 & 0 & 0 & 0.380319 & 0 & 
\multirow{3}{*}{1} & 
\multirow{3}{*}{$0$} & 
\multirow{3}{*}{$9.258201$} & 
\multirow{3}{*}{$-1.224744$} & 
\multirow{3}{*}{$9.315885$} & 
\multirow{3}{*}{$0$} \\
0 & 0 & -0.439155 & 0 & 0 & & & & & & \\
-1 & 0 & -0.358568 & 0 & 0 & & & & & & \\
\hline
1 & 0 & 0.462910 & -0.231455 & 0 & 
\multirow{3}{*}{2} & 
\multirow{3}{*}{$-9.258201$} & 
\multirow{3}{*}{$-3.585685$} & 
\multirow{3}{*}{$-8.660254$} & 
\multirow{3}{*}{$1.463850$} & 
\multirow{3}{*}{$9.710083$} \\
0 & 0 & 0.267261 & 0.267261 & 0 & & & & & & \\
-1 & 0 & -0.377964 & 0 & 0 & & & & & & \\
\hline
1 & 0 & 0.306186 & 0.387298 & 0 & 
\multirow{3}{*}{3} & 
\multirow{3}{*}{$-1.224744$} & 
\multirow{3}{*}{$8.660254$} & 
\multirow{3}{*}{$9.721111$} & 
\multirow{3}{*}{$9.082951$} & 
\multirow{3}{*}{$9.082951$} \\
0 & 0 & 0.111803 & 0.353553 & 0 & & & & & & \\
-1 & 0.499999 & 0.223606 & 0 & 0 & & & & & & \\
\hline
1 & 0 & 0.258774 & -0.146385 & -0.273861 & 
\multirow{3}{*}{4} & 
\multirow{3}{*}{$-9.315885$} & 
\multirow{3}{*}{$1.463850$} & 
\multirow{3}{*}{$-9.082951$} & 
\multirow{3}{*}{$-5.946187$} & 
\multirow{3}{*}{$12.377975$} \\
0 & -0.316227 & 0.464834 & -0.059761 & 0 & & & & & & \\
-1 & -0.387298 & 0.327326 & 0 & 0 & & & & & & \\
\hline
1 & 0.353553 & 0 & -0.073192 & -0.223606 & 
\multirow{3}{*}{5} & 
\multirow{3}{*}{$0$} & 
\multirow{3}{*}{$-9.710083$} & 
\multirow{3}{*}{$9.082951$} & 
\multirow{3}{*}{$-12.377975$} & 
\multirow{3}{*}{$0$} \\
0 & 0.387298 & 0.084515 & -0.223606 & 0 & & & & & & \\
-1 & 0.316227 & 0.267261 & 0 & 0 & & & & & & \\
\end{tabular}
\end{footnotesize}
\end{table}

\begin{table}
\caption{Expansion coefficients and reduced matrix elements for $\so(5)$ irrep 
$(1,{\scriptstyle \frac{3}{2}})$.}
\begin{scriptsize}
\begin{tabular}{@{}c@{}|@{}ccccc@{}|@{}c@{}|cccccccc}
\multicolumn{6}{c}{a-coefficients} & \multicolumn{9}{c}{reduced $\hat{O}$ matrix elements} \\\\
 & \multicolumn{5}{c}{$K-m$} & & & & & & & & & \\
$m$ & $5$ & $2$ & $-1$ & $-4$ & $-7$ & $L$ & $\frac{1}{2}$ & $\frac{3}{2}$ & $\frac{5}{2}$ & ${\frac{7}{2}}_1$ & ${\frac{7}{2}}_2$ & $\frac{9}{2}$ & $\frac{11}{2}$ & $\frac{13}{2}$ \\
\hline \hline
$\frac{3}{2}$ & 0 & 0 & 0.396130 & 0 & 0 & 
\multirow{4}{*}{$\frac{1}{2}$} & 
\multirow{4}{*}{$0$} & 
\multirow{4}{*}{$0$} & 
\multirow{4}{*}{$-8.964214$} & 
\multirow{4}{*}{$-7.598895$} & 
\multirow{4}{*}{$-5.949760$} & 
\multirow{4}{*}{$0$} & 
\multirow{4}{*}{$0$} & 
\multirow{4}{*}{$0$} \\
$\frac{1}{2}$ & 0 & 0 & 0.560213 & 0 & 0 & & & & & & & & & \\
-$\frac{1}{2}$ & 0 & 0 & 0 & 0 & 0 & & & & & & & & & \\
-$\frac{3}{2}$ & 0 & 0.431252 & 0 & 0 & 0 & & & & & & & & & \\
\hline
$\frac{3}{2}$ & 0 & 0 & 0.662853 & 0 & 0 & 
\multirow{4}{*}{$\frac{3}{2}$} & 
\multirow{4}{*}{$0$} & 
\multirow{4}{*}{$-4.225771$} & 
\multirow{4}{*}{$9.486832$} & 
\multirow{4}{*}{$-12.794126$} & 
\multirow{4}{*}{$4.038603$} & 
\multirow{4}{*}{$6.866065$} & 
\multirow{4}{*}{$0$} & 
\multirow{4}{*}{$0$} \\
$\frac{1}{2}$ & 0 & 0 & 0 & 0 & 0 & & & & & & & & & \\
-$\frac{1}{2}$ & 0 & 0.441902 & -0.441902 & 0 & 0 & & & & & & & & & \\
-$\frac{3}{2}$ & 0 & -0.360811 & 0 & 0 & 0 & & & & & & & & & \\
\hline
$\frac{3}{2}$ & 0 & 0 & 0.177154 & -0.560213 & 0 & 
\multirow{4}{*}{$\frac{5}{2}$} & 
\multirow{4}{*}{$-8.964214$} & 
\multirow{4}{*}{$-9.486832$} & 
\multirow{4}{*}{$-7.606388$} & 
\multirow{4}{*}{$-6.510670$} & 
\multirow{4}{*}{$-2.843828$} & 
\multirow{4}{*}{$-5.163977$} & 
\multirow{4}{*}{$12.928374$} & 
\multirow{4}{*}{$0$} \\
$\frac{1}{2}$ & 0 & -0.528173 & 0.417558 & 0 & 0 & & & & & & & & & \\
-$\frac{1}{2}$ & 0 & -0.771447 & 0.192861 & 0 & 0 & & & & & & & & & \\
-$\frac{3}{2}$ & 0 & -0.578585 & 0 & 0 & 0 & & & & & & & & & \\
\hline
$\frac{3}{2}$ & 0 & -0.666136 & 0.063536 & -0.324072 & 0 & 
\multirow{4}{*}{${\frac{7}{2}}_1$} & 
\multirow{4}{*}{$7.598895$} & 
\multirow{4}{*}{$-12.794126$} & 
\multirow{4}{*}{$6.510670$} & 
\multirow{4}{*}{$-0.634212$} & 
\multirow{4}{*}{$-10.096484$} & 
\multirow{4}{*}{$3.705020$} & 
\multirow{4}{*}{$-1.797956$} & 
\multirow{4}{*}{$-13.815832$} \\
$\frac{1}{2}$ & 0 & -0.525280 & 0.059902 & 0.223850 & 0 & & & & & & & & & \\
-$\frac{1}{2}$ & 0 & -0.216048 & -0.432096 & 0 & 0 & & & & & & & & & \\
-$\frac{3}{2}$ & 0.208236 & 0.463617 & 0 & 0 & 0 & & & & & & & & &\\
\hline
$\frac{3}{2}$ & 0 & 0.008606 & 0.559438 & 0.254742 & 0 & 
\multirow{4}{*}{${\frac{7}{2}}_2$} & 
\multirow{4}{*}{$5.949760$} & 
\multirow{4}{*}{$4.038603$} & 
\multirow{4}{*}{$2.843828$} & 
\multirow{4}{*}{$-10.096484$} & 
\multirow{4}{*}{$6.403215$} & 
\multirow{4}{*}{$17.274085$} & 
\multirow{4}{*}{$8.534437$} & 
\multirow{4}{*}{$0.178497$} \\
$\frac{1}{2}$ & 0 & -0.229439 & 0.527444 & 0.309605 & 0 & & & & & & & & & \\
-$\frac{1}{2}$ & 0 & 0.169828 & 0.339656 & 0 & 0 & & & & & & & & & \\
-$\frac{3}{2}$ & -0.724373 & 0.298976 & 0 & 0 & 0 & & & & & & & & & \\
\hline
$\frac{3}{2}$ & 0 & 0.394124 & 0.105334 & -0.394124 & 0 & 
\multirow{4}{*}{$\frac{9}{2}$} & 
\multirow{4}{*}{$0$} & 
\multirow{4}{*}{$-6.866065$} & 
\multirow{4}{*}{$-5.163977$} & 
\multirow{4}{*}{$-3.705020$} & 
\multirow{4}{*}{$-17.274085$} & 
\multirow{4}{*}{$0.530457$} & 
\multirow{4}{*}{$8.616404$} & 
\multirow{4}{*}{$12.924606$} \\
$\frac{1}{2}$ & 0 & 0.371584 & 0.347585 & -0.371584 & 0 & & & & & & & & & \\
-$\frac{1}{2}$ & 0.455096 & 0.198620 & 0.198620 & 0 & 0 & & & & & & & & & \\
-$\frac{3}{2}$ & 0.643602 & 0.344020 & 0 & 0 & 0 & & & & & & & & & \\
\hline
$\frac{3}{2}$ & 0 & 0.226294 & -0.140542 & -0.078390 & 0.335648 & 
\multirow{4}{*}{$\frac{11}{2}$} & 
\multirow{4}{*}{$0$} & 
\multirow{4}{*}{$0$} & 
\multirow{4}{*}{$-12.928374$} & 
\multirow{4}{*}{$-1.797956$} & 
\multirow{4}{*}{$8.534437$} & 
\multirow{4}{*}{$-8.616404$} & 
\multirow{4}{*}{$-11.283690$} & 
\multirow{4}{*}{$13.274930$} \\
$\frac{1}{2}$ & -0.316452 & 0.480398 & -0.265008 & -0.021335 & 0 & & & & & & & & & \\
-$\frac{1}{2}$ & -0.467431 & 0.618721 & -0.149346 & 0 & 0 & & & & & & & & & \\
-$\frac{3}{2}$ & -0.443444 & 0.382507 & 0 & 0 & 0 & & & & & & & & & \\
\hline
$\frac{3}{2}$ & 0.403399 & -0.023853 & -0.029214 & 0.075431 & -0.223765 & 
\multirow{4}{*}{$\frac{13}{2}$} & 
\multirow{4}{*}{$0$} & 
\multirow{4}{*}{$0$} & 
\multirow{4}{*}{$0$} & 
\multirow{4}{*}{$13.815832$} & 
\multirow{4}{*}{$-0.178497$} & 
\multirow{4}{*}{$12.924606$} & 
\multirow{4}{*}{$-13.274930$} & 
\multirow{4}{*}{$-2.884731$} \\
$\frac{1}{2}$ & 0.474678 & 0 & -0.123946 & 0.202403 & 0 & & & & & & & & & \\
-$\frac{1}{2}$ & 0.474678 & 0.116857 & -0.233715 & 0 & 0 & & & & & & & & & \\
-$\frac{3}{2}$ & 0.350573 & 0.286242 & 0 & 0 & 0 & & & & & & & & & \\
\end{tabular}
\end{scriptsize}
\end{table}

\section{Concluding remarks}

As mentioned in the introduction, the group SO(5) and its Lie algebra $\so(5)$ arise
in many physical contexts and, depending on the situation, their irreps are needed in
different bases. 
In particular, they are needed in bases which reduce one of the following:
(i) the U(2) subgroup whose Lie algebra is the grade-conserving intrinsic subalgebra shown
in Fig.\ \ref{fig:weights};
(ii) the SO(4) subgroup generated by the $S$ and $T$ root vectors shown in Fig.\ 
\ref{fig:weights}; and
(iii) the geometric SO(3) subgroup considered in this paper.
VCS theory \cite{VCS} has now been used in systematic ways to construct SO(5) irreps in all
of these bases.

VCS theory was originally formulated to construct the holomorphic representations
of the compact and non-compact discrete-series representations of the symplectic algebras.
A first application constructed the irreps of the non-compact ${\frak sp}(6,\Rb)$
algebra (also called ${\frak sp}(3,\Rb)$)  in a U(3) coupled basis. A next application, by
Hecht and Elliott \cite{HE}, derived similar results for the compact ${\frak sp}$(4) algebra
(isomorphic to $\so$(5)) in a U(2)-coupled basis.

The common feature of the holomorphic VCS representations, which  makes them particularly
simple \cite{Hechtbook}, is their use of  Abelian orbiter groups with
Lie algebras spanned by commuting sets of grade-lowering (or grade-raising) operators to
complement the grade conserving intrinsic subgroups which generate highest-
(or lowest-) grade intrinsic vectors for a VCS representation.
A problem is that Abelian orbiter groups cannot be found in many important situations.
For example, in constructing irreps of $\so$(5) in an $\so$(4)-coupled basis, it is natural
to use SO(4) as the intrinsic subgroup.
However, the complementary $\so$(5) lowering operators (i.e., the operators
$F_-$ and $X_-$ of Fig.\ \ref{fig:weights}) do not then commute.
Fortunately, it was possible to extend the VCS construction \cite{RBH} to accommodate
this situation. 
It was also discovered  \cite{su3so3} that VCS theory could be extended to give the
much needed irreps of, for example, $\su$(3) in an $\so$(3) basis by using SO(3) itself as
an orbiter group to complement an intrinsic U(2) group. This method was subsequently used
\cite{RH} to construct the so-called one-rowed irreps of $\so$(5) (the irreps of type
$(v,0)$ in the notation of this paper) for which only scalar valued VCS wave functions are
needed. 
The techniques have been further developed in this paper to give the generic $\so$(5)
irreps in what is the most sophisticated explicit VCS construction of irreps to date.

Some novel features which considerably simplify the construction of VCS representations
were introduced in Ref.\ \cite{RH} and further developed here.
In particular, we have found that the Hilbert space $\mathcal{H}$ of all SO(5) VCS
wave functions is a model space for SO(5), in the sense that it contains precisely one copy
of every irrep. A particularly valuable feature of this model space  is that
it is a ring in as much as the product of any two VCS wave functions is another VCS wave
function in $\mathcal{H}$. This observation greatly facilitates the construction of an ${\rm
SO(5)}\supset{\rm SO(3)}$-coupled basis for $\mathcal{H}$; one has simply to construct
multiple SO(3)-coupled products of the generating functions given by the VCS wave functions
for the fundamental (1,0) and (0,$\hf$) irreps. 
As a result, it is very easy to construct basis wave functions for any
$\so$(5) irrep. 
It is also shown that, once a basis has been determined, the matrices of
the $\so$(5) Lie algebra in this basis can be determined algebraically, cf. Eqs.\
(\ref{eq:104}) and (\ref{eq:108}).  
These results are remarkable in view of the fact that
there is generally a multiplicity in the ${\rm SO(5)}\supset{\rm SO(3)}$ reduction of the
representation space. 
It must be remembered, however, that the basis in which these results
are obtained is not orthonormal. 

An important component of VCS theory is its incorporation of
algorithms, that go under the name of $K$-matrix theory \cite{Kmatrix}, whose purpose is to
determine the inner product of a Hilbert space,
construct an orthonormal basis, and transform a VCS irrep into a unitary representation
whenever it is equivalent to a unitary irrep (or more generally to an isometric
irrep when it is equivalent to an isometric irrep.) 
For the irreps considered in this paper, a particularly simple version of the
$K$-matrix transformation is given \cite{Kmatrix} by first finding an orthogonal basis in
which off-diagonal matrix elements of the SO(5) octupole operator are zero
between states of the same SO(3) angular momentum.  This is simply achieved by
diagonalisation of sub-blocks of the octupole matrices.  The renormalisation factors needed
to give an orthonormal basis can then be read off as shown in Section \ref{sec:trans}.
Unfortunately, this simple method does not work for states of angular momentum $L=0,\hf,$
or $1$, when there are multiple states of these $L$ values.
One must then resort to less simple $K$-matrix methods.

To summarise, the construction of the SO(3)-reduced matrices of the octupole
operators in a generic  $\so$(5) irrep is  achieved in three simple steps: 
(i) construction of basis wave functions with good ${\rm SO}(3)$ angular momentum quantum
numbers; 
(ii) calculation of the reduced matrix elements of the $\so$(5) octupole operators in this
basis; and \\
(iii) transforming these matrices to those of a unitary representation in an orthonormal
basis.

Step (i) is simply achieved for a $(vf)$ irrep by first computing the basis wave
functions for the $(v0)$ and $(0f)$ irreps and
then taking their SO(3)-coupled products, as indicated in Section \ref{sect:basis}.
Step (ii) is achieved by use of the analytical expressions given in Section
\ref{sect:irreps} in terms of SU(2)  Clebsch-Gordan coefficients  and the coefficients of
the wave functions of step (i). 
Finally step (iii) is achieved by the unitarisation process described above and in Section
\ref{sec:trans}.
The routine has been coded in MAPLE which is amenable to both algebraic and
numeric computations.  
Steps (i) and (ii) are carried out algebraically; thus the results in Tables
\ref{tab:1} - \ref{tab:3} are given in exact arithmetic.
Step (iii), in which diagonalisations are used to orthogonalise states of a common angular
momentum, is done numerically; thus the results given in Section
\ref{sec:results} are floating point numbers.  
The current code has not been designed to handle large-dimensional irreps.  
However, should large  irreps be of interest, the routine could be coded entirely
in a numerical language such as FORTRAN,   MATLAB,  or C$^{++}$.

The representations of $\so(5)$ in an SO(3)-coupled basis are of
interest for several reasons.
The so-called one-row irreps of type $(v0)$ feature in the 
Bohr-Mottelson and IBM-1
collective models and, consequently, have received much attention 
\cite{CMS,RH,RTR}.
The generic irreps, constructed in this paper, show up in these 
collective models whenever
the neutron and proton degrees of freedom are considered independently.
They could also prove useful in the classification of shell-model 
states of fermions in an
angular momentum $l=2$ orbital.
Generic $\so$(5) irreps also show up in supersymmetric boson-fermion 
models
\cite{IachelloDSS,Schmidt}.
For example, the irreps of the orthosymplectic osp(5/4) superalgebra of 
Iachello's model
contain irreps of the $\overline{\so}$(5) subalgebra contained in the 
chain
\be \frak{osp}(5/4) \supset \so(5)\oplus \frak{sp}(4) \supset 
\overline{\frak{sp}}(4)
\simeq \overline{\so}(5) ,\ee
where $\overline{\so}$(5) here signifies the subalgebra of 
$\so(5)\oplus \frak{sp}(4)$
obtained by adding the corresponding infinitesimal generators of the isomorphic 
$\so(5)$ and
$\frak{sp}(4)$ algebras.
The construction of VCS representations of orthosymplectic 
superalgebras was
considered by LeBlanc and Rowe \cite{LeBR} in a natural extension of 
the holomorphic
representations to include representations over Grassmann as well as 
complex variables.
Thus, it would be useful for the development of supersymmetric models 
of coupled
boson-fermion systems to extend the construction given in this paper 
for $\so(5)\simeq
\frak{usp}(4)$ irreps in an SO(3) basis to the irreps of the 
orthosymplectic algebras.

It is interesting to note that, in the present construction of $\so(5)$ 
irreps, the
$\so(5)$ algebra has been realised as a combination of $L=2$ and 
$L=\frac{3}{2}$ boson operators.
The use of bosons of half-odd integer angular momentum is admittedly 
unusual in physics.
But, in spite of the spin-statistics theorem, there is no algebraic
reason forbidding their use in this way.
The fact is that bilinear products of either boson or fermion operators 
of any given angular
momentum obey precisely the same commutation relations.
However, the number of $\so(5)$ irreps that can be built up with 
fermions is severely
limited by the Pauli principle whereas with bosons there is no such 
limitation.

  A major motivation for the present study was that the methods 
developed for $\so(5)$
would serve as prototype examples of what can be done in more general 
situations.
Constructing the irreps of a Lie algbra (or a Lie group) in a basis 
which reduces a
non-canonical subgroup chain is generally much more challenging than 
in a basis which
reduces a canonical subgroup chain.
Thus, it is an order of magnitude simpler to construct $\so(5)$ irreps 
in a canonical ${\rm
SO(5)} \supset  {\rm SO(4)} \supset {\rm SO(3)}$ basis than in the 
geometrical SO(3) basis
considered in this paper. For example, analytical expressions are 
readily derived for the
irreps of $\su(3)$ in the canonical ${\rm SU(3)} \supset {\rm SU(2)}$ 
basis.
But, prior to the development of apposite VCS techniques \cite{su3so3}, 
special ad hoc
methods were needed to obtain these irreps in ${\rm SU(3)} \supset {\rm 
SO(3)}$ bases.
A possible application of the methods developed in
this paper might be to construct irreps of   the exceptional Lie 
algebra $\frak{g}_2$ in an
SU(3) $\supset$ SO(3)-coupled basis.
This would appear to be possible by choosing intrinsic states which 
carry an irrep
of a $\u(2)$ subalgebra, corresponding to a pair of short  $\frak{g}_2$ 
roots, and
employing  SU(3) as the orbiter group.
Construction of such irreps of $\frak{g}_2$ is of physical interest as 
a step towards the
construction of $\so(7)$ irreps in a $G_2 \supset {\rm SO(3)}$-coupled 
basis.
The need for such irreps  would surface if it is were considered 
desirable to formulate
a collective model of octupole  dynamics analogous to Bohr's quadrupole 
model or to
classify shell-model states of fermions in an angular momentum $l=3$ 
orbital.

\section{Acknowledgments}

PST would like to thank the NSERC, the Sumner Foundation and the University of Toronto 
for financial assistance during the course of this research.  DJR and JR also 
thank the NSERC for partial support of their research.  The authors are pleased to 
acknowledge the expert assistance of Santo D'Agostino, Juliana Carvellho, and especially 
Chairul Bahri early in the project.


\begin{thebibliography}{99}
\bibitem{Bahri} 
A MAPLE routine for the $\so(5)\downarrow\so(3)$ branching rules was provided to us by C.\ Bahri.
\end{thebibliography}
\end{document}